\documentclass[11pt,twoside]{article}
\usepackage{amsmath,epsfig,amssymb,amsfonts,amsthm,epsfig,verbatim}
\usepackage{color,graphicx,lscape,longtable}
\usepackage{mathrsfs}
\usepackage{amsbsy}
\usepackage{multicol}
\usepackage{dsfont}
\usepackage{indentfirst}%
\usepackage{booktabs}
\usepackage[figuresright]{rotating}
\usepackage{caption}
\usepackage{float}
\usepackage{multirow} 
\usepackage{setspace} 
\usepackage{fancyhdr} 
\usepackage{calligra,algorithm,algorithmicx,algpseudocode}
\usepackage{mathtools}
\usepackage{hyperref}
\hypersetup{
    colorlinks=true, 
    linkcolor=blue,  
    citecolor=blue,  
    urlcolor=blue,   
}

\usepackage{color}

\usepackage[authoryear, sort]{natbib}
\usepackage{enumitem}
\usepackage{datetime}   

\allowdisplaybreaks

\usepackage[margin=1.0in]{geometry}
\makeatletter
\def\singlespace{\def\baselinestretch{1}\@normalsize}

\renewcommand{\hat}{\widehat}

\newtheorem{thm}{\underline{\bf Theorem}}

\newtheorem{lem}{\underline{\bf Lemma}}
\newtheorem{corollary}{\underline{\bf Corollary}}

\newtheorem{remark}{\underline{\bf Remark}}
\newtheorem{assume}{\underline{\bf Assumption}}

\newcommand{\sumt}{\ensuremath{\sum_{t=1}^{T}}}

\allowdisplaybreaks
\newcommand{\cd}{\stackrel{d}{\rightarrow}}

\def\mR{\mathbb{R}}

\def\argmin{\mbox{argmin}}

\def\0{{\bf 0}}

\def\B{{\bf B}}

\def\v{{\boldsymbol v}}

\def\X{{\bf X}}

\def\I{{\bf I}}

\def\Z{{\bf Z}}

\def\bbeta{{\boldsymbol{\beta}}}
\def\btheta{{\boldsymbol \theta}}
\def\bdelta{{\boldsymbol \delta}}

\def\balpha{{\boldsymbol \alpha}}

\def\bSigma{{\bf \Sigma}}

\def\mF{\mathcal{F}}
\def\mI{\mathcal{I}}
\def\mE{\mathbb{E}}

\def\boxit#1{\vbox{\hrule\hbox{\vrule\kern6pt\vbox{\kern6pt#1\kern6pt}\kern6pt\vrule}\hrule}}

\LARGE
\def\title.arg{
	\LARGE{\bf Sparse Interval-valued Time Series Modeling with Machine Learning}
	\footnote{Corresponding author: Yuying Sun (sunyuying@amss.ac.cn), Academy of Mathematics and Systems Science, Chinese Academy of Sciences, 55 Zhongguancun East Road, Haidian District, Beijing 100190, China. All remaining errors are solely ours. We thank seminar participants in the 40th International Symposium on Forecasting and the 10th International Symposium of Econometric Analysis and Forecasting at Dalian for their comments and suggestions. Bao thanks support from China NNSF Grant No. 72403232. Sun thanks support from China NNSF Grant (Nos. 72322016, 72073126, 72091212, 71973116) and Young Elite Scientists Sponsorship Program by CAST (YESS) [2020QNRC001]. Hong and Wang thank support from China NNSF Fundamental Scientific Center Project (No.71988101) entitled as Econometric Modeling and Economic Policy Studies.}
	}

\def\author.arg{
Haowen Bao$^{a,b}$, Yongmiao Hong$^{a,b}$, Yuying Sun$^{a,b, \dag}$, Shouyang Wang$^{a,b}$\\
\small $^a$Academy of Mathematics and Systems Science, Chinese Academy of Sciences, China\\
\small $^b$School of Economics and Management, and MOE Social Science Laboratory of Digital Economic\\\small Forecasts and Policy Simulation, University of Chinese Academy of Sciences, Beijing, China}

\def\tit.arg{\LARGE
{Penalized Interval Linear Regression Models}
}

\def\key.arg{{
 Forecasting; High-dimensional modeling; Interval-valued time series; Machine learning; Sparse linear regression}
}

\def\abst.arg{
\baselineskip=24pt
{
By treating intervals as inseparable sets, this paper proposes sparse machine learning regressions for high-dimensional interval-valued time series. With LASSO or adaptive LASSO techniques, we develop a penalized minimum distance estimation, which covers point-based estimators are special cases. We establish the consistency and oracle properties of the proposed penalized estimator, regardless of whether the number of predictors is diverging with the sample size. Monte Carlo simulations demonstrate the favorable finite sample properties of the proposed estimation. Empirical applications to interval-valued crude oil price forecasting and sparse index-tracking portfolio construction illustrate the robustness and effectiveness of our method against competing approaches, including random forest and multilayer perceptron for interval-valued data. Our findings highlight the potential of machine learning techniques in interval-valued time series analysis, offering new insights for financial forecasting and portfolio management.}
}

\def\date.arg{\large {November 7, 2024}}

\numberwithin{equation}{section}
\begin{document}
\begin{center}
	\begin{spacing}{2.0}
		{\Large \title.arg}
	
		\author.arg

		\date.arg
	\end{spacing}
\end{center}
\vspace{-0.8cm}

\centerline{\bf Abstract}
\abst.arg

\pagestyle{plain}

\newpage
\section{Introduction} \label{chap:introduction}
\setcounter{equation}{0}
\baselineskip=24pt
Machine learning techniques have revolutionized data analysis across various domains,  providing powerful tools for forecasting economic variables and financial outcomes \citep{gu2020empirical,babii2022machine,bia2024double}. However, their
application to interval-valued data remains largely unexplored, despite the increasing prevalence of such data in economic and financial spheres. interval-valued data, or more generally symbolic data, offer richer information (e.g., trends and volatility) than point-valued data within the same time period, and avoid noise contained in high-frequency data  \citep{bock2000symbolic, billard2002symbolic, bock1999analysis, golan2017interval}. In the data-rich environment, interval-valued data are prevalent across diverse situations, including interval-valued salary in the job advertisements \citep{zhong2023feature}, high-low asset returns \citep{Gloria2013Constrained, wang2012cipca}, a minimax regret portfolio selection problem \citep{GIOVE2006253}, high-low livestock prices \citep{2016Interval, Zhang2020hybrid}, minimum-maximum daily air quality data \citep{YANG2019336}, and variation of Cholesterol level \citep{wang2012linear}. These interval-valued data, coupled with advanced machine learning techniques, offers an opportunity to enhance the analysis and forecasting of complex phenomena. As far as we know, there is little work on developing spares regression models for high-dimensional interval-valued data.

Our attempt in this article is to investigate the crucial problem of interval-valued machine learning methods in high-dimensional contexts, for which we immediately face several unique theoretical challenges in contrast to the case of point-valued time series data. 
First, interval-valued data introduce inherent complexities due to their distinct algebraic properties and operational rules, necessitating novel approaches beyond standard point-valued techniques. Second, different from quadratic loss with $L_1$ penalties for the existing literature for point-valued machine learning methods, we need to develop a proper loss function to simultaneously capture potential information contained in interval-valued data and yield sparse regression model. Third,  the diverging dimensional setting in this paper substantially complicates the theoretical derivations, particularly in establishing the consistency and asymptotic normality of our estimators. A key reason for this is that classical large sample theories, such as the law of large numbers and central limit theorem, developed for point-valued data, cannot be directly applied to interval-valued data.

To address these challenges,  we propose a sparse linear regression via machine learning tools to select relevant interval-valued features and optimize model parameters by treating intervals as inseparable sets. 
The estimation procedure is developed within a penalized minimum distance framework based on $D_K$-metrics with adaptive LASSO type penalty, and we introduce a novel interval-based least angle regression (ILARS) algorithm to solve the resulting optimization problem. Leveraging central limit theories based on $D_K$-distance, we establish asymptotic properties of the proposed parameter estimators. Specifically, under suitable regularity conditions, we prove that the penalized minimum distance estimators demonstrate consistency and asymptotic normality in both fixed and diverging dimensional settings, achieving the oracle properties\footnote{The oracle properties indicate that the estimators perform as well as if the true submodel were known a priori \citep{2001Variable}}.
In addition, we introduce the interval-based framework of nonnegative garrote regression and ridge regression as additional regularized approaches for interval modeling. Furthermore, when interval-valued data reduces to point-valued data, the proposed loss function with a suitable kernel function is equivalent the quadratic loss. Simulation studies demonstrate the favorable finite sample properties of the proposed estimation. Empirical applications to  crude oil prices forecasting and  S\&P 100 index-tracking highlight the merits of the proposed method compared to other competing approaches, including  random forest and multilayer perceptron (MLP).

It is worth discussing some key references and outlining our contributions in relation to the most relevant literature. First, in contrast to existing interval regression methods \citep{Gloria2013Constrained,2016Interval,Lin2019extreme,LIMANETO20081500,LIMANETO2010333}, our proposed approach treats intervals as inseparable entities to capture the potential information contained within the intervals. Previous methods, such as that of \cite{Gloria2013Constrained}, construct interval models by modeling upper and lower bounds. While this approach facilitates the direct application of traditional point-data modeling techniques, including maximum likelihood and least squares estimation, it only utilizes the boundary information of intervals, thereby neglecting the potentially valuable information contained within the intervals. Although \cite{Gloria2013Constrained}'s method has been widely adopted, this limitation of considering only boundary points is also prevalent in subsequent studies \citep{DIAS20171118,buansing2020information}. To overcome this methodological shortcoming, our proposed penalized $D_K$-distance measure utilizes the information of boundaries as well as the interior points of an interval. By utilizing the potential information contained in intervals, our approach is expected to achieve superior forecasting accuracy.

Second, the proposed parsimonious model with adaptive LASSO represents the first attempt to simultaneously achieve estimation and consistent variable selection for high-dimensional ITS. While parsimonious regression techniques for point-valued data have been extensively studied and advanced over the past decades, from the seminal least absolute shrinkage and selection operator \citep[LASSO,][]{tibshirani1996regression} to various extensions including adaptive LASSO \citep{zou2006adaptive}, group LASSO \citep{Meier2008grouplasso}, with continuing developments in recent years \citep{bia2024double,gao2024robust,caner2024should}. However, analogous methodologies for interval-valued data remain largely unexplored. While some machine learning approaches for interval-valued data exist, such as artificial neural networks \citep{YANG2019336}, support vector machines \citep{utkin2019imprecise}, and visualization techniques \citep{zhang2022visualization}, none of these methods is designed to handle the dual challenges of high dimensionality and sparsity inherent in modern interval-valued datasets. To the best of our knowledge, only \cite{zhong2023feature} explicitly addresses high-dimensional interval-valued data. Nevertheless, their approach is restricted to cases where only the response variable is interval-valued while predictors remain point-valued, making it unsuitable for our context where both predictors and response are interval-valued.

Third, we establish the consistency and oracle properties of the proposed penalized estimator, regardless of whether the number of predictors is diverging with the sample size. While several methods treating interval variables as inseparable entities exist, their theoretical frameworks are primarily restricted to low dimensions. For instance, \cite{han2016vector} introduced an interval-valued vector autoregressive moving average (IVARMA) model, where each component can be viewed as an autoregressive conditional interval model with interval-valued exogenous variables (ACIX). To capture nonlinear features of ITS, \cite{sun2018threshold} developed threshold autoregressive interval models, which was further extended by \cite{yang2024LSTIAR} to a logistic smooth transition interval autoregressive model. Similar set-based approaches have been widely used in various fields, including stock market \citep{wang2016set}, foreign exchange market \citep{sun2020assessing}, and commodity market \citep{wu2023pass,he2021forecasting}, among many others. However, none of these existing methodologies addresses the challenges posed by high-dimensional interval-valued data, which is the focus of our study.

The rest of this paper is organized as follows. Section 2 introduces the interval-based machine learning regression via adaptive LASSO.
Section 3 describes the proposed estimators' asymptotic properties and introduces the interval-based nonnegative garrote and ridge regression. Section 4 develops the associated asymptotic properties of diverging-dimensional interval regression. Section 5 presents the simulation studies to show the finite sample properties of our method. Section 6 shows two empirical applications on the interval-valued crude oil price forecasting and the interval-based index tracking. Section 7 concludes the paper and discusses its future prospects. The appendix presents technical assumptions, implementation algorithms, and mathematical proofs, with additional simulation results in the supplementary materials.

\section{ Machine learning regression for ITS} \label{sec 2}
We first review the ACIX models in the existing literature \citep{sun2018threshold,he2021forecasting}.
Suppose $\{Y_t\}$ and $\{X_{j,t}\}$ are stationary ITS\footnote{The interval variable is defined as a measurable map on a probability space $(\Omega,\mF,P)$, namely $Y: \Omega \rightarrow I_{\mathbb{R}}$, where $I_{\mathbb{R}}$ is the set of all pairs of ordered numbers in $\mathbb{R}$. Specifically, for any $w$ in $\Omega$, the term of $I_{\mathbb{R}}$ takes the form of $Y(w)=[Y_L(w),Y_R(w)]$, where $Y_R<Y_L$ is allowed. See more discussions in Remark \ref{rem1}}.
Then, an ACIX model\footnote{For any given intervals $Y_1=[Y_{1L},Y_{1R}]$, $Y_2=[Y_{2L},Y_{2R}]$ and scalar $c$, the operation rules of intervals are defined as follows: (1) addition: $Y_1+Y_2=[Y_{1L}+Y_{2L},Y_{1R}+Y_{2R}]$; (2) Hukuhara's difference: $Y_1-Y_2=[Y_{1L}-Y_{2L},Y_{1R}-Y_{2R}]$; (3) scalar multiplication: $c\cdot Y_1=[c\cdot Y_{1L},c\cdot Y_{1R}]$.}
can be expressed as
\begin{align} \label{eq-2.1}
	Y_t=\alpha_0+\beta_0 I_0+\sum_{j=1}^{q}\beta_j Y_{t-j}+\sum_{j=0}^{s}\bdelta'_j \X_{t-j}+u_t, \quad t=1,...,T,
\end{align}
where $\X_t=(X_{1,t},...,X_{J,t})'$, $\alpha_0,\beta_j$ and $\bdelta_j=(\delta_{j,1},\dots,\delta_{j,J})'$ are unknown scalar parameters, $I_0$ is the interval unit element $[-\frac{1}{2},\frac{1}{2}]$, and $u_t$ is an interval martingale difference sequence (IMDS), satisfying $\mathbb{E}[u_t|\mI_{t-1}]=[0,0]$ with $\mI_{t-1}$ being the information set. Let $\btheta=(\alpha_0,\beta_0,\beta_1,\dots,\beta_q,\bdelta_0',\dots,\bdelta_s')'=(\theta_1,\dots,\theta_{p})'$ and $\Z_t=([1,1],I_0,Y_{t-1},\dots,Y_{t-q},\X_t',\dots,\X_{t-s}')'$, where $p$ is the dimension of all parameters being either fixed or diverging as $T\rightarrow\infty$. More discussions of ACIX model\footnote{
Similar to the point-valued case, we consider the interval regression model \eqref{eq-2.1} to be correctly specified in conditional mean, if there exists a true parameter $\btheta^0 \in \mR^p$ such that $\mE[Y_t|\Z_t] = \Z_t'\btheta^0$. Otherwise, we can define the pseudo-true parameter $\btheta^*$ as $\btheta^*=\argmin \mE[\lVert Y_t-\Z_t'\btheta\rVert _K^2]$, where $\Z_t'\btheta^*$ represents the optimal linear combination in the sense of minimizing the $D_K$ distance. Moreover, when \eqref{eq-2.1} contains no exogenous variables, it reduces to the ACI model.}
and the definition of intervals can be found in the existing literature, e.g., \cite{he2021forecasting}, \cite{Yang2016}, and among others.

The ACIX model serves as a generalization of the popular ARX-type model, commonly employed for point-valued time series analysis. It provides a framework to capture the temporal dependence of interval processes observed in economics and finance, such as volatility or range clustering and level effects. However, in the era of big data, the dimension of $\Z_t$ is often sufficiently large to encompass the underlying structure of high-dimensional interval-valued data, and there may exist sparsity within the predictors. For instance, when modeling the dynamics of interval-valued stock returns influenced by multiple factors \citep{guo2023statistical}, such as supply, demand, geopolitical tensions, and technological advancements, or when analyzing interval-valued macroeconomic indicators using various predictors \citep{koop2023bayesian}, like GDP growth or inflation forecasting ranges subject to uncertainties arising from consumer confidence, trade policies, and industrial productivity. Additionally, this challenge arises in modeling interval-valued exchange rates  \citep{premanode2013improving}, which are impacted by numerous factors, including economic fundamentals, market sentiment, and central bank interventions. In such scenarios, the ACIX model becomes less suitable due to its fixed number of parameters and inability to handle redundant variables. 

\subsection{Penalized minimum $D_K$-distance estimation}

To select important interval-valued predictors from a large dataset, we propose a penalized minimum distance estimation for the ACIX model. { Without loss of generality, we assume that the response and covariates are standardized.} Our objective is to estimate the unknown regression coefficients by solving the following penalized regression problem based on $D_K$-distance:  
\begin{align} \label{eq2.2}
\hat{\btheta}_T=\argmin_{\btheta}\sumt\lVert Y_t-\Z_t'\btheta\rVert _K^2+\lambda_T\sum_{j=1}^p {w}_j\lvert \theta_j\rvert,
\end{align}
where $\lVert\cdot\rVert_K$ is the $D_K$ norm derived from the $D_K$-distance, $\lambda_T$ is the tuning parameter, $w_j$ is a known adaptive weight for $j=1,\dots,p$, and $T$ is the sample size. In practice, we can set the adaptive weight as $\hat{w}_j=1/|\tilde{\theta}_j|^\gamma$, where $\tilde{\btheta}=(\tilde{\theta}_1,...,\tilde{\theta}_p)'$ is the minimum $D_K$-distance estimator\footnote{
The minimum $D_K$-distance estimator $\tilde{\btheta}=\argmin \sumt\lVert Y_t-\Z_t'\btheta\rVert _K^2$, and it has been proven to be a $\sqrt{T}$ consistent estimator with fixed dimension $p$.} \citep{han2016vector}, and $\gamma$ is a given constant.

The $D_K$-distance is a metric for measuring the distance between two interval-valued variables \citep{korner2002variance,han2016vector,sun2018threshold,he2021forecasting}. Specifically, the $D_K$-distance and its derived norm $\lVert\cdot\rVert_K$ are defined as:
\begin{align} \label{eq2.3}
\lVert Y_t-\Z_t'\btheta\rVert _K^2 &= D_K^2(Y_t,\Z_t'\btheta)=\int_{(u,v)\in S^0}[s_{Y_t}(u)-s_{\Z_t'\btheta}(u)][s_{Y_t}(v)-s_{\Z_t'\btheta}(v)]dK(u,v),
\end{align}
where $K(u,v)$ is a symmetric positive definite kernel function for $u, v \in S^0=\{1,-1\}$, and $s_{Y_t}(u)$ is the support function of interval $Y_t$ defined on the unit sphere $S^0=\{-1,1\}$ as:
\[
s_{Y_t}(u)=\left\{
    \begin{aligned}
    \sup_{y\in Y_t}\{u \cdot y\} \text{, when } Y_{L,t}\leq Y_{R,t} \\
    \inf_{y\in Y_t}\{u \cdot y\} \text{, when } Y_{R,t}\leq Y_{L,t},
    \end{aligned}
            \right.
\]for $u\in S^0$.
It is noteworthy that the operation rules and $D_K$ norm facilitate the construction of a complete normed linear space. Furthermore, we can endow this space with an inner product induced by the $D_K$-distance metric, denoted as $\langle\cdot,\cdot\rangle_K$.\footnote{
For example, suppose $X_1$ and $X_2$ are two intervals. The inner product of them is $\langle s_{X_1},s_{X_2} \rangle_K$. For simplicity of notation, we extend the use of $\langle\cdot,\cdot\rangle_K$ to also denote the multiplication of interval matrices, which can be obtained by replacing the pointed-valued multiplication with inner product for intervals.}

\begin{remark} \label{rem1}
The intervals $Y_t$ and $X_{j,t}$ in \eqref{eq-2.1} are extended intervals, a concept introduced by \cite{Kaucher1980}. Mathematically, extended intervals generalize classical intervals by allowing the left bound to exceed the right bound, providing an algebraically closed space where operations and proofs can be constructed in closed form. Combined with interval operation rules, particularly Hukuhara's difference \citep{1967Integration}, they enable the construction of a complete normed linear space through the $D_K$ norm.
From a practical perspective, this framework has been widely adopted in interval analysis \citep{dimitrova1992extended, benhamou2006continuous, sun2018threshold, sahu2024efficient}, as it naturally accommodates real-world scenarios where such ordering may occur. For instance, in analyzing household income differentials, the wife's income (right bound) may exceed the husband's income (left bound).
\end{remark}

\subsection{Discussion of special cases}
Our penalized minimum $D_K$-distance estimation covers several classical special cases of interval models. The $D_K$-distance is, to a certain degree, equivalent to the $d_W$ distance\footnote{
The $d_W$ distance for intervals is defined as $d_W(A,B)=\sqrt{\int_{[0,1]}(A(\omega)-B(\omega))^2dW(\omega)}$ for $A,B\in I_\mR$, where $W(\omega)$ is a probability measure on the real Borel space $([0,1],\B([0,1]))$.} introduced by \cite{bertoluzza1995new}, with the advantage of being more computationally tractable.
The $d_W$ distance measure involves not only distances between extreme points with weights $W(0)$ and $W(1)$, but also distances between interior points in the intervals with weights $W(\omega),0<\omega<1$. It is interesting to see that the $D_K$ metric, as a equivalence of the $d_W$ metric, preserves this property, which is demonstrated through examples in the special cases.
In the following, we investigate various special choices of kernel $K(u,v)$ and discuss the corresponding penalized regression. For notational convenience, we denote the kernel $K$ more concisely as: $K(1,1)=a$, $K(1,-1)=K(-1,1)=b$, and $K(-1,-1)=c$. Let $Y_{m,t}=(Y_{L,t}+Y_{R,t})/2$, $Y_{r,t}=Y_{R,t}-Y_{L,t}$, $\Z_{m,t}=(\Z_{L,t}+\Z_{R,t})/2$, $\Z_{r,t}=\Z_{R,t}-\Z_{L,t}$ be the midpoints and ranges of $Y_t$ and $\Z_t$. Then, denote $\Delta_{m,t}=Y_{m,t}-\Z_{m,t}'\btheta$, $\Delta_{r,t}=Y_{r,t}-\Z_{r,t}'\btheta$, $\Delta_{L,t}=Y_{L,t}-\Z_{L,t}'\btheta$, and $\Delta_{R,t}=Y_{R,t}-\Z_{R,t}'\btheta$.

{\bf Case 1.} $a=1/4, b=-1/4, c=1/4$. The $D_K$ norm becomes $||Y_t-\Z_t'\bbeta||_K^2=D_K(Y_t,\Z_t'\bbeta)^2 = \Delta_{m,t}^2$.\footnote{In this case, the $D_K$-distance is  equivalent to the $d_W$ distance, where $W(\omega)$ is a distribution such that $W(1/2)=1$ and $0$ otherwise.} 
Thus, the penalized minimum distance estimation is obtained by
\begin{eqnarray} \label{case1}
    \hat{\btheta}_T=\argmin_\btheta \sumt (Y_{m,t}-\Z_{m,t}'\btheta)^2 + \lambda_T\sum_{j=1}^p \frac{1}{|\tilde{\theta}_j|^\gamma}\lvert \theta_j\rvert,
\end{eqnarray}
which is the adaptive LASSO estimation for the midpoints of intervals. Under this kernel function, our interval model effectively utilizes only the midpoint information of the intervals. Especially, if there is no penalty in \eqref{case1}, note that our method degenerates to the midpoints method proposed by \cite{billard2000regression}.

{\bf Case 2.} $a=1, b=1, c=1$. The $D_K$ norm becomes $||Y_t-\Z_t'\bbeta||_K^2=D_K(Y_t,\Z_t'\bbeta)^2 = \Delta_{r,t}^2$. This leads to the following optimization problem:
\[
\hat{\btheta}_T=\argmin_\btheta \sumt (Y_{r,t}-\Z_{r,t}'\btheta)^2 + \lambda_T\sum_{j=1}^p \frac{1}{|\tilde{\theta}_j|^\gamma}\lvert \theta_j\rvert.
\]
In this case, our method only use the range information of the ITS. It is equivalent to the adaptive LASSO estimation for the ranges of intervals. 

{\bf Case 3.} $a,c>0, b=0$. We have $||Y_t-\Z_t'\bbeta||_K^2=D_K(Y_t,\Z_t'\bbeta)^2=a\Delta_{R,t}^2+c\Delta_{L,t}^2$.\footnote{If $a+c=1$, the choice of such a kernel $K$ is equivalent to the choice of $W(\omega)$ in $d_W$ distance with $W(\omega)$ follows a Bernoulli distribution with $W(0)=c$ and $W(1)=a$.}
The penalized estimation \eqref{eq2.2} becomes
\begin{eqnarray} \label{case3}
    \hat{\btheta}_T=\argmin_\btheta \sumt \bigg[a(Y_{R,t}-\Z_{R,t}'\btheta)^2+c(Y_{L,t}-\Z_{L,t}'\btheta)^2\bigg] + \lambda_T\sum_{j=1}^p \frac{1}{|\tilde{\theta}_j|^\gamma}\lvert \theta_j\rvert.
\end{eqnarray}
In this case, the estimator $\hat{\btheta}_T$ is obtained by minimizing the square errors of weighted bounds with a penalization. Moreover, when there is no penalty (i.e., $\lambda_T=0$), our method encompasses several popular special cases. First, note that \eqref{case3} is similar to the constrained Minmax method. The Minmax method estimates the lower and upper bounds of the intervals using different parameter vectors \citep{billard2002symbolic}, thereby ignoring the dependence between the bounds. Then, \eqref{case3} can also be seen as the bivariate regression in \cite{brito2007modelling} with a constraint. With this case, our model is essentially equivalent to using information about the interval's left and right bounds.

{\bf Case 4.} $a=c, |b|<a$. It follows that, $||Y_t-\Z_t'\bbeta||_K^2 = D_K(Y_t,\Z_t'\bbeta)^2 = \frac{a+b}{2}\Delta_{r,t}^2+2(a-b)\Delta_{m,t}^2$.\footnote{If $a-b=1$ and $b\leq0$, the choice of such a kernel $K$ is equivalent to the choice of $W(\omega)$ in $d_W$ distance with $W(\omega)$ follows a distribution such that $W(0)=a+b$, $W(1/2)=-4b$, and $W(1)=c+b$.}
Then, \eqref{eq2.2} takes the form of
\begin{eqnarray} \label{case4}
    \hat{\btheta}_T=\argmin_\btheta \sumt \bigg[\frac{a+b}{2}(Y_{r,t}-\Z_{r,t}'\btheta)^2+2(a-b)(Y_{m,t}-\Z_{m,t}'\btheta)^2\bigg] + \lambda_T\sum_{j=1}^p \frac{1}{|\tilde{\theta}_j|^\gamma}\lvert \theta_j\rvert.
\end{eqnarray}
In this case, the estimator $\hat{\btheta}_T$ is obtained by the penalized square errors of weighted ranges and midpoints. When $\lambda_T=0$, equation \eqref{case4} provides an approach similar to the well-known CRM method proposed by \cite{LIMANETO20081500}, but with additional constraints. Importantly, our method under this kernel function is equivalent to utilizing both midpoint and range information of intervals.

{\bf Case 5.} $a\neq c, b\neq 0$. We have, $||Y_t-\Z_t'\bbeta||_K^2 = D_K(Y_t,\Z_t'\bbeta)^2 = a\Delta_{R,t}^2 + c\Delta_{L,t}^2 -2b\Delta_{R,t}\Delta_{L,t}$ or equivalently $||Y_t-\Z_t'\bbeta||_K^2=(a+2b+c)/4\Delta_{r,t}^2 + (a-2b+c)\Delta_{m,t}^2+(a-c)\Delta_{r,t}\Delta_{m,t}$.\footnote{if $a+c-2b=1$ and $b<0$, the $D_K$-distance is equivalent to $d_W$ distance with distributions as $W(0)=a+b$, $W(1/2)=-4b$, $W(1)=c+b$, and 0 otherwise.}
In this case, $\hat{\btheta}_T$ is obtained by solving the following optimization problem:
\begin{eqnarray} \label{case5}
    \hat{\btheta}_T=\argmin_\btheta \sumt \bigg[ a\Delta_{R,t}^2 + c\Delta_{L,t}^2 -2b\Delta_{R,t}\Delta_{L,t} \bigg] + \lambda_T\sum_{j=1}^p \frac{1}{|\tilde{\theta}_j|^\gamma}\lvert \theta_j\rvert,
\end{eqnarray}
Here, \eqref{case5} could capture the information in $\Delta_{R,t}$, $\Delta_{L,t}$, and $\Delta_{R,t}\Delta_{L,t}$. Utilizing the cross product information will enhance estimation efficiency.

\section{Asymptotic properties}

\subsection{Consistency and oracle properties}
In this section, we examine the asymptotic properties of estimation under the condition that the dimension of the predictors in the penalized ACIX model is large but fixed. In the following, the $L_2$ norm of any vector is denoted by $||\cdot||$. Theorems \ref{thm1} and \ref{thm2} establish the consistency and asymptotic normality, respectively, of the penalized estimators for interval linear regression. The necessary conditions for these theorems are listed in Appendix A.

\begin{thm} \label{thm1}
	Given Assumptions \ref{ass1}-\ref{ass4} and \ref{ass8}, the penalized minimum distance estimator in \eqref{eq2.2} is consistent, that is, $||\hat{\btheta}_T-\btheta^0||=O_p(1/\sqrt{T})$, as $T\rightarrow \infty$.
\end{thm}

Theorem \ref{thm1} specifically states that under the imposed assumptions, the distance between the estimated parameter vector $\hat{\btheta}_T$ and the true parameter vector $\btheta^0$ in the penalized interval regression converges in probability to zero at a rate of $O_p(1/\sqrt{T})$. This theorem provides a theoretical guarantee for the consistency of the estimator $\hat{\btheta}_T$, ensuring that it converges to the true parameter values at a well-defined rate as the sample size $T$ increases.

In the following analysis, we establish two main asymptotic results: the consistency of interval-valued variable selection and the asymptotic normality of non-zero coefficient estimators. For exposition purposes, we assume without loss of generality that the first $k_0$ predictors are the true variables with non-zero coefficients, while the remaining $m_0=p-k_0$ predictors are redundant with zero coefficients. Let $\btheta^0=(\btheta_1^{0'},\btheta_2^{0'})'$, where $\btheta_1^0$ is a $k_0\times 1$ vector of non-zero coefficients and $\btheta^0_2$ is a $m_0\times1$ vector of zero coefficients, i.e., $\btheta^0_1\neq\0$ and $\btheta^0_2=\0$. Note that this partition is solely for theoretical analysis, as the true non-zero and zero coefficients are unknown in practice. Let $\hat{\btheta}_T=(\hat{\btheta}_{1T}',\hat{\btheta}_{2T}')'$ denote the estimator corresponding to $\btheta^0_1$ and $\btheta^0_2$, respectively. Further, let $\Z_{1t} = (Z_{1t},...,Z_{k_0t})'$ represent the vector of the first $k_0$ covariates.

\begin{thm}[Oracle properties] \label{thm2}		
	Under Assumptions \ref{ass1}-\ref{ass4} and \ref{ass8}, the penalized interval regression estimation via adaptive LASSO satisfies the following properties:
		
	(i) Consistency in variable selection: $\lim_{T\rightarrow\infty}P(\hat{\btheta}_{2T}=\0)=1$;
	
	(ii) Asymptotic normality: $\sqrt{T}(\hat{\btheta}_{1T}-\btheta_{1}^0)\stackrel{d}\rightarrow N(\0,C_{11}^{-1} \mE[\langle s_{\Z_{1t}},s_{u_t}\rangle_K \langle s_{u_t},s'_{\Z_{1t}}\rangle_K] C_{11}^{-1})$, as $T\rightarrow\infty$, where $C_{11}$ is a $k_0\times k_0$ block of matrix $\mE[\langle s_{\Z_t},s'_{\Z_t}\rangle_K]$
	\footnote{ In \cite{han2012autoregressive}, some asymptotic properties are derived as
	$\frac{1}{T}\sumt\langle s_{\Z_t},s'_{\Z_t}\rangle_K\stackrel{p}\rightarrow \mE[\langle s_{\Z_t},s'_{\Z_t}\rangle_K]$ and
	$\sqrt{T}(\tilde\btheta-\btheta^0)\stackrel{d}\rightarrow N(0,[\mE\langle s_{\Z_t},s'_{\Z_t}\rangle_K]^{-1} \mE[\langle s_{\Z_t},s_{u_t}\rangle_K \langle s_{u_t},s'_{\Z_t}\rangle_K][\mE\langle s_{\Z_t},s'_{\Z_t}\rangle_K]^{-1})$. It is important to note that the asymptotic variance in Theorem \ref{thm2} cannot be simplified under conditional homoskedasticity, see more details in Theorem 3.2 of \cite{han2012autoregressive}.}
corresponding to the nonzero coefficients.
\end{thm}

Theorem \ref{thm2} states two asymptotic properties of the penalized interval regression. The first property is the consistency in variable selection, meaning that the probability of correctly identifying the zero coefficients approaches 1 as the sample size increases. The other property is that estimated coefficients of active variables follow an asymptotic normal distribution with mean zero and variance, in terms of $C_{11}^{-1} \mE[\langle s_{\Z_{1t}},s_{u_t}\rangle_K \langle s_{u_t},s'_{\Z_{1t}}\rangle_K] C_{11}^{-1}$.

\subsection{Properties of other interval-based regularized methods}
Following the spirit of \cite{breiman1995better}, we propose the interval-based nonnegative garrote, a machine learning technique for adaptive feature selection. We also analyze the relationship between penalized interval regression via adaptive LASSO and via nonnegative garrote. The nonnegative garrote for intervals is equivalent to minimizing the following loss function,
\begin{align} \label{eq garrote}
	\sumt \lVert Y_t-\sum_{j=1}^p Z_{jt}\tilde{\theta}_jc_j\rVert _K^2+\lambda_T\sum_{j=1}^p c_j, \quad \text{subject to} \quad \forall j, c_j\geq 0,
\end{align}
where $c_j$ is a constant, $\lambda_T$ is the tuning parameter, $\hat{\theta}^j_{garrote}= c_j\tilde{\theta}_j$ is the interval-based nonnegative garrote estimator, and $\tilde{\theta}_j$ denotes the minimum $D_K$-distance estimator as defined in Section 2.1. Based on equation \eqref{eq garrote} and Theorem \ref{thm2}, we can derive the following corollaries.

\begin{corollary} \label{cor 1}
	The nonnegative garrote for interval-valued data is a special case of the penalized interval regression via adaptive LASSO with $\gamma=1$, $\hat{w}_j=1/|\tilde{\theta}_j|$ and $\theta_j\tilde{\theta}_j\geq 0$ for any $j=1,...,m$.
\end{corollary}

\begin{corollary} \label{cor 2}
	Suppose that the conditions in Theorem \ref{thm2} hold. Then the interval-based nonnegative garrote is consistent for variable selection.
\end{corollary}

If the conditions in Corollary \ref{cor 1} are satisfied, i.e., $\gamma$ in estimation \eqref{eq2.2} is set to be 1 and $\hat{w}_j=1/|\tilde{\theta}_j|$, the penalized estimation for interval regression takes the form:
\begin{align} \label{eq garotte1}
	\hat{\btheta}_T=\argmin_{\btheta}\sumt\lVert Y_t-\Z_t'\btheta\rVert _K^2+\lambda_T\sum_{j=1}^p \frac{\lvert \theta_j\rvert}{|\tilde{\theta}_j|^\gamma}.
\end{align}
Let $\theta_j=c_j \tilde{\theta}_j$ and $\theta_j/\tilde{\theta}_j\geq 0$ (or $\theta_j \tilde{\theta}_j\geq 0$). Then, \eqref{eq garotte1} is equivalent to \eqref{eq garrote}. Corollary \ref{cor 2} shows the consistency of variable selection of the interval-based nonnegative garrote.

Next, we propose a ridge regression method for interval-valued data, introducing a regularized approach that directly incorporates interval structures. The interval-based ridge regression is given by
\begin{align} \label{IB ridge}
		\hat{\btheta}_{ridge}=\argmin_{\btheta} \sumt\lVert Y_t-\Z_t'\btheta\rVert _K^2 +\lambda_T\sum_{j=1}^p \theta_j^2.
\end{align}
Since the optimization \eqref{IB ridge} is a quadratic problem, the following corollary presents a closed-form solution for parameter estimation derived through direct differentiation.

\begin{corollary} \label{lem ridge}
	With Assumptions \ref{ass1}-\ref{ass4}, the solution of nonlinear quadratic convex programming problem \eqref{IB ridge} can be expressed as:
	 \begin{align} \label{eq ridge est}
	 	\hat{\btheta}_{ridge}=(\sumt\langle s_{\Z_t},s'_{\Z_t}\rangle_K+\lambda_T \I)^{-1}(\sumt\langle s_{\Z_t},s_{Y_t}\rangle_K),
	 \end{align}
	 where $\lambda_T$ is the tuning parameter and $\I$ is the identity matrix.
\end{corollary}

\section{Asymptotic properties in diverging dimensions
}
In the preceding sections, we have analyzed the asymptotic properties of the interval linear regression model via adaptive LASSO when the dimension of regressors is fixed. Besides, the number of regressors could be diverging, which has been studied for point-valued data, like \cite{fan2004nonconcave} and \cite{huang2008asymptotic}. In this section, we focus on a diverging dimension of regressors as the sample size increases, that is, $p$ grows to infinity at some slower rates than the sample size $T$. We first show the consistency of the minimum $D_K$-distance estimator for ACIX model when the dimension of predictors is diverging.

\begin{thm} \label{thm6}
	Given Assumptions \ref{ass1}-\ref{ass5}, the minimum $D_K$-distance estimator of \eqref{eq-2.1} $\tilde{\btheta}_T$ is $\rho_{1T}^{-1}\sqrt{p/T}$ consistent, namely
	\begin{eqnarray}
		||\tilde{\btheta}_T-\btheta^0||=O_p(\rho_{1T}^{-1}\sqrt{p/T}), 
	\end{eqnarray}
 as $T\rightarrow\infty$, where $\rho_{1T}$ is the smallest eigenvalue of $\bSigma_T=\frac{1}{T}\sumt\langle s_{\Z_t},s_{\Z_t'}\rangle_K$.
\end{thm}
Theorem \ref{thm6} is a generalization of estimation consistency of the conventional ACIX model. It can also be seen as a generalization of the ordinary least square (OLS) estimation consistency with diverging dimension for the interval-valued case.

Without loss of generality, we assume that the coefficients of the first $k_T$ variables are non-zero. It follows that the coefficients of the last $m_T=p-k_T$ variables are zeros. Moreover, we still let $\Z_{1t}$ denote the first $k_T$ covariates and $\bSigma_{1T} = \frac{1}{T}\sumt\langle s_{\Z_{1t}},s_{\Z_{1t}'}\rangle_K$. Before giving the consistency and oracle properties of the penalized estimation with diverging dimension, we first show the following Lemma.

\begin{lem} \label{lemA1}
	Let $\v$ be a $p\times 1$ vector. For any positive $\delta$, under Assumption \ref{ass1},
	\begin{eqnarray*}
		\mE \sup_{||\v||<\delta} |\sumt\langle s_{u_t},s_{\Z_t'\v}\rangle_K|\leq \delta\sigma\sqrt{Tp}.
	\end{eqnarray*}
\end{lem}

\begin{thm} \label{thm7}
	Let $\hat{\btheta}_T$ be the estimator of the interval-based adaptive LASSO regression. Suppose Assumptions \ref{ass1}-\ref{ass5}, \ref{ass9}(a) and \ref{ass10} hold. Let $h_T=\rho_{1T}^{-1}\sqrt{p/T}$ and $h_T'=[(p+\lambda_Tk_T(\rho_{1T}^{-1}(p/T)^{1/2}+1)^{-\gamma})/(T\rho_{1T})]^{1/2}$. Then $\hat{\btheta}_T$ is consistent and $||\hat{\btheta}_T-\btheta^0||=O_p(\min\{h_T,h_T'\})$, as $T\rightarrow\infty$.
\end{thm}
Note that $\rho_{1T}$ appears in the denominators of $h_T$ and $h_T'$, which makes it possible that $h_T'$ may converge to zero faster than $h_T$ if $\rho_{1T}\rightarrow0$. Additionally, if we suppose that there exists a positive constant $\rho_1$ and $\rho_2$ such that $0<\rho_1<\rho_{1T}<\rho_2<\infty$, Theorem \ref{thm7} yields that the convergence rate of $h_T=O_p(\sqrt{p/T})$ and $h_T\leq h_T'$. Thus we have $||\hat{\btheta}_T-\btheta^0||=O_p(\sqrt{p/T})$. This condition is also common in the existing  literature, such as Condition (F) in \cite{fan2004nonconcave}. Furthermore, if $p$ is finite, Theorem \ref{thm7} degenerates to Theorem \ref{thm1}.

\begin{thm} \label{thm8}
	Let $\hat{\btheta}_T=(\hat{\btheta}_{1T}',\hat{\btheta}_{2T}')'$ be the solution of \eqref{eq2.2}, where $\hat{\btheta}_{1T}$ and $\hat{\btheta}_{2T}$ are estimators of $\btheta_{1}^0$ and $\btheta_2^0$, respectively. Suppose Assumptions \ref{ass1}-\ref{ass7}, \ref{ass9} and \ref{ass10} are satisfied. We have the following properties:

	(i) Consistency in variable selection: $\lim_{T\rightarrow\infty}P(\hat{\btheta}_{2T}=\0)=1$.

	(ii) Let $\xi_T^2=\sigma^2\balpha_T'\Psi\balpha_T$, where $\Psi=\frac{1}{T}\sumt\mE(\bSigma_{1T}^{-1}\langle s_{\Z_{1t}},s_{u_t}\rangle_K\langle s_{\Z_{1t}'},s_{u_t}\rangle_K\bSigma_{1T}^{-1})$, $\balpha_T$ is any $k_T\times1$ vector satisfying $||\balpha_T||=1$. Then, as $T\rightarrow\infty$, we have
	\begin{align*}
		T^{1/2}\xi_T^{-1}\balpha_T'(\hat{\btheta}_{1T}-\btheta_1^0)
		=T^{1/2}\xi_T^{-1}\sumt\balpha_T'\bSigma_{1T}^{-1}\langle s_{\Z_{1t}},s_{u_t}\rangle_K+o_p(1) \cd N(0,1).
	\end{align*}
\end{thm}

Theorem \ref{thm8} (i) indicates that the penalized linear regression for interval-valued data is consistent in variable selection, that is, the estimators of the zero coefficients are exactly zero with high probability when $T$ is large. Moreover, Theorem \ref{thm8} (ii) states that the estimators of the nonzero parameters have the asymptotic normal distribution when the number of parameters diverges. Similar results in the point-valued case can be found in existing literature, such as \cite{fan2004nonconcave}, \cite{huang2008asymptotic} and \cite{huang2008adaptive}. While these studies considered the independent and identically distributed point-valued random variables, this paper proves the oracle properties for martingale difference ITS. Furthermore, Theorem \ref{thm8} demonstrates that our proposed penalized model can effectively identify the true non-zero coefficients while shrinking irrelevant coefficients to zero, thus achieving model sparsity. This variable selection property is particularly valuable in high-dimensional interval-valued settings, where it not only enhances model interpretability but also potentially improves prediction accuracy by reducing model complexity.

\begin{remark}
	The condition $p<T$ is necessary for the identification and consistent estimation of the linear regression model for interval-valued data. This condition is frequently satisfied in economic and financial applications, which justifies its inclusion in our model assumptions. However, when $p>T$, the minimum $D_K$-distance estimators are no longer feasible as initial estimators for the adaptive weights. In analogous point-valued models with i.i.d. errors, \cite{huang2008adaptive} demonstrated that marginal regression estimators are zero-consistent under a partial orthogonality condition. However, the proof of estimators' asymptotic consistency requires a commonly used proposition in \cite{van1997weak}, which cannot be applied anymore for ITS. The exploration of asymptotic properties for penalized linear interval regression estimators in ultrahigh-dimensional scenarios is a subject left for future research endeavors.
\end{remark}

\begin{remark}
 In the context of diverging number of regressors, the nonnegative garrote method can also be extended to the interval-valued case. It is easy to verify that Corollary \ref{cor 1} and \ref{cor 2} hold in this case. This extension provides an alternative approach to variable selection and parameter estimation for high-dimensional interval data. Besides, the interval-based ridge regression can be solved by \eqref{eq ridge est}. Additionally, Corollary \ref{lem ridge} can also be obtained.
\end{remark}

\section{Simulation} \label{sec sim}
This section investigates the finite sample performance of the proposed penalized estimation for interval regression. The two-stage minimum $D_K$-distance estimators are considered as estimated adaptive weights of the penalized estimation \citep{he2021forecasting, han2016vector}, and the tuning parameter $\lambda_T$ is selected by a five-fold cross-validation process.

\subsection{ACI-based data generating processes} \label{Sec: Simu A}
First, we consider the data generating process (DGP) as follows:
\begin{align} \label{DGP 1}
	Y_t=\alpha_0 +\beta_0 I_0 +\sum_{j=1}^{p-2}\delta_j X_{j,t} +u_t, \quad t=1,\dots,T,
\end{align}
where $Y_t$, $X_{j,t}$, and $u_t$ are all interval variables, $\btheta=(\alpha_0,\beta_0,\delta_1,\dots,\delta_{p-2})'$ is the given point-valued coefficients. The ordered pairs $X_{j,t}=(X_{L,j,t},X_{R,j,t}), j=1,...,p-2,$ are generated from the bivariate normal distributions with non-zero covariance matrix. To generate the interval innovations $\{u_t\}$, we employ an ACI(1,0) process following \cite{sun2018threshold}:
\begin{align} \label{u_t DGP}
	Y_t=\alpha_0+\beta_0 I_0+\beta_1 Y_{t-1}+u_t,
\end{align}
where the parameters $(\alpha_0, \beta_0, \beta_1)'$ are estimated by a two-stage minimum $D_K$-distance method; $Y_t=\ln(P_t)-\ln(P_{t-1})$ and $P_t$ are designated as the time series of the daily S\&P 500 index data for the period January 2, 2015 to December 31, 2019, with its bounds as the high and low prices in day $t$. From model \eqref{u_t DGP}, we have the estimated interval innovation, namely $\hat{u}_t=Y_t-(\hat{\alpha}_0+\hat{\beta}_0 I_0+\hat{\beta}_1 Y_{t-1})$. We then generate $\{u_t\}$ via the naive bootstrapping from $\{\hat{u}_t\}$, with sample size $T$. In addition, following \cite{han2016vector} and \cite{sun2018threshold}, a two-stage minimum $D_K$-distance method is also used here with some given preliminary kernels $K$ in the first stage. We take a preliminary kernel in the first step with $K=(a,b,c)=(5,1,1)$.\footnote{
\cite{han2012autoregressive} proved that the two-stage $D_K$-distance estimator is asymptotically most efficient among all symmetric positive definite kernels satisfying $K(1,1)>0$, $K(1,1)K(-1,-1)>K(1,-1)^2$ and $K(1,-1)=K(-1,1)$. Thus, as sample size $T$ increases to infinity, the choice of kernel $K$ in the first stage has little impact on the optimal kernel derived by the two-stage estimation.}

In our simulation experiments, we consider two DGPs:

{\bf DGP 1:} The dimension of predictors is fixed at $p=10$. Following \cite{zou2006adaptive}, the initial coefficients are set as $\btheta=(\alpha_0,\beta_0,\delta_1,\dots,\delta_8)'=(0,0,3,1.5,0,0,2,0,0,0)'$. We examine the estimators under sample sizes $T=20, 40, 80$.

{\bf DGP 2:} The dimension of predictors diverges as the sample size increases. We set $p=[3T^{1/3}]$, where $[\cdot]$ represents the largest integer not exceeding $3T^{1/3}$. The initial coefficients are $\btheta=(0,0,11/4,-23/6,37/12,-13/9,1/3,0,...,0)'$. We consider sample sizes $T=100,200,400,800$.

Each experiment is repeated 1000 times. To evaluate the performance of our estimation method, we employ bias (Bias), standard deviation (SD), and root mean square error (RMSE) for each estimated parameter $\hat{\theta}_j$, i.e.,
\begin{align}
	&\text{Bias}(\hat{\theta}_j)=\frac{1}{N}\sum_{i=1}^{N}(\hat{\theta}_j^{(i)}-\theta_j), \label{eq Bias}\\
	&\text{SD}(\hat{\theta}_j)=[\frac{1}{N}\sum_{i=1}^{N}(\hat{\theta}_j^{(i)}-\bar{\theta}_j)^2]^{\frac{1}{2}},\label{eq SD} \\
	&\text{RMSE}(\hat{\theta}_j)=[\frac{1}{N}\sum_{i=1}^{N}(\hat{\theta}_j^{(i)}-\theta_j)^2]^{\frac{1}{2}}, \label{eq RMSE}
\end{align}	
where $N=1000$ is the number of replications, $\theta_j$ is the true parameter value, and $\bar{\theta}_j=\frac{1}{N}\sum_{i=1}^N\hat{\theta}_j^{(i)}$ is the average of the estimated $\hat{\theta}_j$ across all replications.
	
The LARS algorithm has been used to compute the solution path for the LASSO problem in point-valued cases \citep{Efron2004Least,Tibshirani2013The}. For our interval-valued case, we propose an interval-based LARS algorithm, as detailed in Appendix B, to obtain the estimators. We set the parameter $\gamma$ to 0.5 and 1 in our experiments.

\subsection{Bivariate normal distribution}
In this section, the interval innovation $(u_{L,t},u_{R,t})$ is generated from a bivariate normal distribution, $(u_{L,t},u_{R,t}) \sim i.i.d. N(0,\bSigma^0)$. The DGP for this case can be expressed as:
\begin{align} \label{DGP:bivariate}
\begin{aligned}
		Y_{L,t} = \alpha_0-\frac{1}{2}\beta_0+\sum_{j=1}^{p-2}\delta_jX_{L,j,t}+u_{L,t}\\
		Y_{R,t} = \alpha_0+\frac{1}{2}\beta_0+\sum_{j=1}^{p-2}\delta_jX_{R,j,t}+u_{R,t}	
\end{aligned}
\end{align}
where $\alpha_0$, $\beta_0$, and $\delta_j$ are given initial scalar parameters, and $\bSigma^0$ is a $2\times 2$ positive definite matrix with diagonal elements equal to 1 and all other elements equal to 0.75. The regressors $(X_{L,j,t},X_{R,j,t})$ $(j=1,\ldots,p-2; t=1,\ldots,T)$ are also generated from bivariate normal distributions with non-zero covariance matrices.

In this section, we also consider two types of DGPs: (1) where $p$ is fixed, and (2) where $p$ diverges as the sample size increases. We define these DGPs as follows:

{\bf DGP 3:} $Y_t=[Y_{L,t},Y_{R,t}]$ and $u_t=[u_{L,t},u_{R,t}]$ are generated according to \eqref{DGP:bivariate}. We set $p=10$, $\btheta=(\alpha_0,\beta_0,\delta_1,\dots,\delta_8)'=(0,0,3,1.5,0,0,2,0,0,0)'$, and $T=20, 40, 80$.

{\bf DGP 4:} $Y_t=[Y_{L,t},Y_{R,t}]$ and $u_t=[u_{L,t},u_{R,t}]$ are generated according to \eqref{DGP:bivariate}. We set $p=[3T^{1/3}]$, $\btheta=(0,0,11/4,-23/6,37/12,-13/9,1/3,0,\ldots,0)'$, and $T=100,200,400,800$.

All other parameter settings remain the same as those in Section \ref{Sec: Simu A}. We employ the criteria defined in equations \eqref{eq Bias} - \eqref{eq RMSE} to evaluate the performance of our estimators.

\subsection{Evaluation results}
Panel A of Table \ref{tab bias} reports the Bias, SD and RMSE of our method and the minimum $D_K$-distance estimators of ACIX model based on DGP 1, with $\gamma=0.5$. Several observations emerge from this panel. First, the Bias, SD and RMSE of each estimator whose true value is set to be zero are approaching zero as the sample size $T$ increases, consistent with asymptotic efficiency of variable selection in Theorem \ref{thm2}. For example, RMSE of $\alpha_0$ decreases from $3.1521 \times 10^{-3}$ to $0.4728\times 10^{-3}$ as $T$ increases from 20 to 80. Second, for the nonvanishing coefficients, the evaluation criteria of estimators of $\delta_1,\delta_2$ and $\delta_5$ converge to zero as $T$ increases. These observations indicate the consistency of the estimated nonvanishing parameters and provide a finite sample evidence for the oracle properties. For example, SD of $\delta_1$ decreases from $0.4867\times 10^{-3}$ to $0.1588\times 10^{-3}$ as $T$ increases from 20 to 80. Third, our method yields substantially improved estimates compared to the minimum $D_K$-distance method. As evidenced in Panel A, our approach demonstrates superior performance relative to ACIX, producing Bias, SD, and RMSE values that more closely approximate zero. This underscores the enhanced efficacy of our penalized method over the minimum $D_K$-distance estimation in scenarios characterized by model sparsity. For example, when the sample size $T=80$, the Bias for $\delta_3$ of our method is $0.0111\times 10^{-3}$, which is smaller than $0.0354\times 10^{-3}$ of the minimum $D_K$-distance method.
When $u_t$ follows a bivariate normal distribution, similar results can be obtained from Panel B of Table \ref{tab bias}.
\begin{table}[htbp]
	\scriptsize
	\centering
	\caption{The Bias, SD, and RMSE of the estimated parameters.}
	\begin{tabular}{llcccccccccc}
	\toprule
	& & $\alpha_0$ & $\beta_0$ & $\delta_1$ & $\delta_2$ & $\delta_3$ & $\delta_4$ & $\delta_5$ & $\delta_6$ & $\delta_7$ & $\delta_8$\\
	\midrule
	\multicolumn{2}{l}{\emph{Panel A}}&\\
	\cline{1-2}
	T=20\\
	Bias	&	PLR	&	-0.0012 &	0.1585 	&	-0.0722 &	-0.0341 &	0.0302 	&	0.0114 	&	-0.0481 &	0.0281 	&	-0.0224 &	-0.0002 	\\
			&	ACIX	&	-1.0764 &	-1.3270 &	0.0278 	&	0.0367 	&	0.0499 	&	0.0367 	&	-0.0545 &	0.0507 	&	0.0543 	&	0.0041 	\\
	SD		&	PLR	&	3.1680 	&	0.8209 	&	0.4867 	&	0.5585 	&	0.2915 	&	0.2343 	&	0.3580 	&	0.3183 	&	0.2712 	&	0.2282 	\\
			&	ACIX	&	9.6777 	&	9.1656 	&	0.8833 	&	0.7050 	&	0.9143 	&	0.5730 	&	0.5573 	&	0.6937 	&	0.5984 	&	0.6455 	\\
	RMSE	&	PLR	&	3.1521 	&	0.8320 	&	0.4896 	&	0.5567 	&	0.2916 	&	0.2334 	&	0.3594 	&	0.3179 	&	0.2708 	&	0.2270 	\\
			&	ACIX	&	9.7373 	&	9.2612 	&	0.8838 	&	0.7059 	&	0.9156 	&	0.5742 	&	0.5600 	&	0.6956 	&	0.6009 	&	0.6456 	\\
	\midrule
	T=40\\																							
		Bias&	PLR	&	0.0582 	&	0.0037 	&	0.0347 	&	-0.0125 &	-0.0020 &	-0.0080 &	-0.0666 &	-0.0046 &	-0.0171 &	0.0118 	\\
			&	ACIX	&	0.1496 	&	-0.7165 &	0.0087 	&	-0.0065 &	0.0660 	&	0.0076 	&	-0.0180 &	0.0526 	&	-0.0051 &	0.0236 	\\
		SD	&	PLR	&	0.8540 	&	0.7069 	&	0.2570 	&	0.3420 	&	0.1643 	&	0.0851 	&	0.1940 	&	0.1572 	&	0.2075 	&	0.1612 	\\
			&	ACIX	&	3.7575 	&	4.9646 	&	0.3348 	&	0.3815 	&	0.3619 	&	0.3144 	&	0.2880 	&	0.2910 	&	0.4563 	&	0.3209 	\\
		RMSE&	PLR	&	0.8517 	&	0.7034 	&	0.2580 	&	0.3405 	&	0.1635 	&	0.0850 	&	0.2042 	&	0.1565 	&	0.2072 	&	0.1608 	\\
			&	ACIX	&	3.7605 	&	5.0160 	&	0.3350 	&	0.3816 	&	0.3678 	&	0.3144 	&	0.2886 	&	0.2957 	&	0.4563 	&	0.3217 	\\
	\midrule
	T=80\\																							
		Bias&	PLR	&	-0.0407 &	-0.2761 &	0.0410 	&	0.0188 	&	0.0111 	&	-0.0078 &	0.0013 	&	0.0002 	&	-0.0002 &	0.0028 	\\
			&	ACIX	&	0.0195 	&	-0.8104 &	0.0447 	&	0.0309 	&	0.0354 	&	0.0169 	&	0.0276 	&	-0.0155 &	0.0161 	&	0.0161 	\\
		SD	&	PLR	&	0.4734 	&	1.5215 	&	0.1588 	&	0.1991 	&	0.1211 	&	0.0876 	&	0.1695 	&	0.1223 	&	0.1236 	&	0.1052 	\\
			&	ACIX	&	3.2108 	&	3.9481 	&	0.2659 	&	0.2391 	&	0.2679 	&	0.2342 	&	0.2484 	&	0.2344 	&	0.2604 	&	0.2372 	\\
		RMSE&	PLR	&	0.4728 	&	1.5388 	&	0.1632 	&	0.1990 	&	0.1210 	&	0.0875 	&	0.1687 	&	0.1217 	&	0.1229 	&	0.1047 	\\
			&	ACIX	&	3.2109 	&	4.0304 	&	0.2696 	&	0.2411 	&	0.2703 	&	0.2348 	&	0.2499 	&	0.2349 	&	0.2609 	&	0.2378 	\\
	\midrule
	\multicolumn{2}{l}{\emph{Panel B}}&\\
	\cline{1-2}
	T=20\\
		Bias&	PLR	&	0.0027 	&	0.0264 	&	-0.0033 &	-0.0232 &	-0.0035 &	0.0007 	&	-0.0145 &	0.0052 	&	0.0031 	&	-0.0060 	\\
			&	ACIX	&	0.0034 	&	0.0632 	&	-0.0083 &	-0.0007 &	-0.0007 &	-0.0014 &	-0.0014 &	0.0059 	&	-0.0038 &	0.0002 	\\
		SD	&	PLR	&	0.1935 	&	0.2469 	&	0.0733 	&	0.1056 	&	0.0453 	&	0.0289 	&	0.0596 	&	0.0525 	&	0.0405 	&	0.0373 	\\
			&	ACIX	&	1.0113 	&	1.5442 	&	0.0939 	&	0.1142 	&	0.1016 	&	0.0877 	&	0.1046 	&	0.0980 	&	0.0970 	&	0.0849 	\\
		RMSE&	PLR	&	0.1926 	&	0.2471 	&	0.0730 	&	0.1076 	&	0.0452 	&	0.0288 	&	0.0610 	&	0.0525 	&	0.0405 	&	0.0376 	\\
			&	ACIX	&	1.0113 	&	1.5455 	&	0.0943 	&	0.1142 	&	0.1016 	&	0.0877 	&	0.1046 	&	0.0982 	&	0.0971 	&	0.0849 	\\
	\midrule	
	T=40\\																							
		Bias&	PLR	&	-0.0089 &	-0.0350 &	-0.0031 &	-0.0012 &	-0.0010 &	0.0043 	&	-0.0070 &	-0.0005 &	-0.0004 &	0.0022 	\\
			&	ACIX	&	-0.0757 &	-0.1771 &	-0.0007 &	0.0064 	&	0.0065 	&	0.0103 	&	-0.0038 &	0.0016 	&	-0.0014 &	0.0010 	\\
		SD	&	PLR	&	0.3045 	&	0.2670 	&	0.0428 	&	0.0637 	&	0.0248 	&	0.0197 	&	0.0403 	&	0.0242 	&	0.0334 	&	0.0367 	\\
			&	ACIX	&	0.8164 	&	0.9504 	&	0.0565 	&	0.0681 	&	0.0706 	&	0.0548 	&	0.0612 	&	0.0454 	&	0.0558 	&	0.0644 	\\
		RMSE&	PLR	&	0.3032 	&	0.2679 	&	0.0427 	&	0.0634 	&	0.0247 	&	0.0201 	&	0.0406 	&	0.0241 	&	0.0333 	&	0.0366 	\\
			&	ACIX	&	0.8199 	&	0.9668 	&	0.0565 	&	0.0684 	&	0.0709 	&	0.0558 	&	0.0613 	&	0.0454 	&	0.0558 	&	0.0644 	\\
	\midrule
	T=80\\																							
		Bias&	PLR	&	-0.0372 &	0.0301 	&	-0.0033 &	-0.0052 &	0.0008 	&	0.0009 	&	-0.0108 &	-0.0011 &	-0.0017 &	0.0015 	\\
			&	ACIX	&	-0.0933 &	0.0762 	&	0.0020 	&	0.0012 	&	-0.0013 &	-0.0033 &	-0.0101 &	-0.0006 &	-0.0053 &	0.0009 	\\
		SD	&	PLR	&	0.1756 	&	0.1834 	&	0.0303 	&	0.0467 	&	0.0232 	&	0.0154 	&	0.0321 	&	0.0205 	&	0.0208 	&	0.0181 	\\
			&	ACIX	&	0.4591 	&	0.7654 	&	0.0381 	&	0.0544 	&	0.0470 	&	0.0421 	&	0.0455 	&	0.0351 	&	0.0438 	&	0.0423 	\\
		RMSE&	PLR	&	0.1787 	&	0.1850 	&	0.0304 	&	0.0468 	&	0.0231 	&	0.0154 	&	0.0337 	&	0.0205 	&	0.0208 	&	0.0181 	\\
			&	ACIX	&	0.4685 	&	0.7691 	&	0.0381 	&	0.0544 	&	0.0470 	&	0.0423 	&	0.0466 	&	0.0351 	&	0.0441 	&	0.0423 	\\
	\bottomrule
	\end{tabular}
	\caption*{\footnotesize Note: Our method is denoted as PLR, and the minimum $D_k$-distance estimation of ACIX is denoted as ACIX. In panel A, the interval innovation error is generated by ACI model. Additionally, all the values in this panel are obtained by multiplying the original value by $10^3$. In panel B, the interval innovation error is generated by bivariate normal distribution. In addition, $(a,b,c)=(5,1,1)$ and $\gamma=0.5$.}
\label{tab bias}
\end{table}

Table \ref{tab_DGP2} shows the evaluation results of DGP 2 and DGP 4 with $K=(5,1,1)$ and $\gamma=1$. In these cases, our proposed penalized minimum distance estimation still outperforms the minimum $D_K$-distance estimation of ACIX model. In Table \ref{tab_DGP2}, most results of Bias from our estimation provide values closer to zero than those of minimum $D_K$-distance estimation. Nearly all results of SD and RMSE from our estimation are smaller than those of the minimum $D_K$-distance estimation. For example, when $T=400$ and $u_t$ is generated by an ACI process, the Bias, SD, and RMSE of $\delta_1$ are 0.0074, 0.2311, and 0.2312 (all $\times10^{-3}$), which are smaller than those of the benchmark estimation: 0.0079, 0.2312, and 0.2537 (all $\times10^{-3}$). Furthermore, Table \ref{tab_DGP2} shows that our model makes more accurate estimation of zero coefficients. For example, when $T=400$ and $u_t$ is generated by a bivariate normal distribution, the evaluation results of $\delta_6$ are -0.0003, 0.0109, and 0.0109, which are 70.0\%, 36.6\%, and 36.6\% better than those of benchmark estimation. A possible explanation for this is that our estimation process could shrink the estimators of the zero coefficients to zero. The evaluation results of other parameters are listed in the online appendices of this paper.
\begin{table}[htbp]
	\scriptsize
	\centering
	\caption{The evaluation results with diverging dimension and $T=100, 200, 400$.}
	\begin{tabular}{lcccccc|cccccc}
	\toprule
	&\multicolumn{2}{c}{Bias}& \multicolumn{2}{c}{SD}& \multicolumn{2}{c}{RMSE} &\multicolumn{2}{|c}{Bias}& \multicolumn{2}{c}{SD}& \multicolumn{2}{c}{RMSE} \\
	& PLR & ACIX & PLR & ACIX &PLR & ACIX & PLR & ACIX & PLR & ACIX &PLR & ACIX \\
	\midrule
	\multicolumn{7}{l}{Panel A1: $T=100$} & \multicolumn{6}{|l}{Panel A2: $T=100$} \\
	$\alpha_0$	&	0.0948 	&	0.4100 	&	0.3911 	&	1.1277 	&	0.4025 	&	1.1999 	&	0.0006 	&	0.0000 	&	0.0469 	&	0.0902 	&	0.0469 	&	0.0902 	\\
	$\beta_0$	&	0.0050 	&	-0.0331 	&	0.5958 	&	1.9965 	&	0.5958 	&	1.9968 	&	-0.0101 	&	0.0061 	&	0.1093 	&	0.2287 	&	0.1098 	&	0.2287 	\\
	$\delta_1$	&	-0.0308 	&	-0.0109 	&	0.5339 	&	0.5768 	&	0.5348 	&	0.5769 	&	-0.0018 	&	-0.0018 	&	0.0348 	&	0.0371 	&	0.0348 	&	0.0371 	\\
	$\delta_2$	&	0.0149 	&	0.0100 	&	0.5319 	&	0.5559 	&	0.5321 	&	0.5560 	&	0.0023 	&	0.0005 	&	0.0449 	&	0.0467 	&	0.0450 	&	0.0467 	\\
	$\delta_3$	&	-0.0008 	&	-0.0009 	&	0.4603 	&	0.4817 	&	0.4603 	&	0.4817 	&	0.0076 	&	0.0056 	&	0.0429 	&	0.0467 	&	0.0436 	&	0.0470 	\\
	$\delta_4$	&	-0.0323 	&	-0.0335 	&	0.5183 	&	0.5239 	&	0.5193 	&	0.5249 	&	0.0014 	&	-0.0010 	&	0.0413 	&	0.0430 	&	0.0413 	&	0.0430 	\\
	$\delta_5$	&	0.0536 	&	0.0631 	&	0.5441 	&	0.5833 	&	0.5468 	&	0.5867 	&	-0.0069 	&	0.0037 	&	0.0375 	&	0.0373 	&	0.0381 	&	0.0374 	\\
	$\delta_6$	&	-0.0200 	&	-0.0153 	&	0.2517 	&	0.6281 	&	0.2525 	&	0.6283 	&	-0.0005 	&	0.0004 	&	0.0158 	&	0.0377 	&	0.0158 	&	0.0377 	\\
	$\delta_7$	&	-0.0084 	&	-0.0920 	&	0.2074 	&	0.4572 	&	0.2076 	&	0.4663 	&	0.0003 	&	-0.0002 	&	0.0197 	&	0.0390 	&	0.0198 	&	0.0390 	\\
	$\delta_8$	&	0.0092 	&	0.0047 	&	0.2829 	&	0.5869 	&	0.2831 	&	0.5869 	&	-0.0027 	&	-0.0025 	&	0.0203 	&	0.0420 	&	0.0205 	&	0.0421 	\\
	$\delta_9$	&	0.0353 	&	0.0000 	&	0.2425 	&	0.5930 	&	0.2451 	&	0.5930 	&	-0.0007 	&	0.0034 	&	0.0161 	&	0.0376 	&	0.0161 	&	0.0378 	\\
	$\delta_{10}$	&	0.0026 	&	0.0016 	&	0.2366 	&	0.5320 	&	0.2366 	&	0.5320 	&	0.0005 	&	-0.0025 	&	0.0139 	&	0.0326 	&	0.0139 	&	0.0327 	\\
	$\delta_{11}$	&	0.0210 	&	0.0184 	&	0.1733 	&	0.4700 	&	0.1746 	&	0.4704 	&	0.0018 	&	-0.0020 	&	0.0243 	&	0.0438 	&	0.0244 	&	0.0439 	\\	
	\midrule
	\multicolumn{7}{l}{Panel B1: $T=200$}& \multicolumn{6}{|l}{Panel B2: $T=200$} \\
	$\alpha_0$	&	0.1598 	&	0.3947 	&	0.4032 	&	0.8348 	&	0.4337 	&	0.9234 	&	0.0019 	&	-0.0010 	&	0.0264 	&	0.0556 	&	0.0265 	&	0.0556 	\\
	$\beta_0$	&	-0.0463 	&	-0.1734 	&	0.4465 	&	1.4727 	&	0.4489 	&	1.4829 	&	0.0023 	&	0.0076 	&	0.0955 	&	0.1878 	&	0.0956 	&	0.1880 	\\
	$\delta_1$	&	0.0123 	&	-0.0246 	&	0.3559 	&	0.3919 	&	0.3561 	&	0.3927 	&	-0.0013 	&	-0.0004 	&	0.0269 	&	0.0281 	&	0.0269 	&	0.0281 	\\
	$\delta_2$	&	0.0084 	&	-0.0114 	&	0.2967 	&	0.3272 	&	0.2968 	&	0.3274 	&	-0.0003 	&	-0.0007 	&	0.0294 	&	0.0306 	&	0.0294 	&	0.0306 	\\
	$\delta_3$	&	0.0304 	&	0.0174 	&	0.3647 	&	0.3781 	&	0.3659 	&	0.3785 	&	-0.0037 	&	-0.0034 	&	0.0283 	&	0.0315 	&	0.0285 	&	0.0317 	\\
	$\delta_4$	&	0.0229 	&	0.0091 	&	0.3506 	&	0.3582 	&	0.3514 	&	0.3583 	&	0.0035 	&	0.0025 	&	0.0210 	&	0.0237 	&	0.0213 	&	0.0238 	\\
	$\delta_5$	&	-0.0117 	&	-0.0366 	&	0.3364 	&	0.3414 	&	0.3366 	&	0.3433 	&	0.0000 	&	0.0053 	&	0.0260 	&	0.0254 	&	0.0260 	&	0.0259 	\\
	$\delta_6$	&	-0.0100 	&	-0.0104 	&	0.1475 	&	0.3751 	&	0.1479 	&	0.3752 	&	0.0025 	&	0.0011 	&	0.0158 	&	0.0271 	&	0.0160 	&	0.0271 	\\
	$\delta_7$	&	0.0325 	&	0.0354 	&	0.1649 	&	0.3669 	&	0.1681 	&	0.3686 	&	0.0028 	&	0.0006 	&	0.0163 	&	0.0308 	&	0.0165 	&	0.0308 	\\
	$\delta_8$	&	0.0031 	&	0.0247 	&	0.1349 	&	0.3361 	&	0.1350 	&	0.3370 	&	-0.0018 	&	0.0049 	&	0.0116 	&	0.0262 	&	0.0117 	&	0.0267 	\\
	$\delta_9$	&	-0.0026 	&	0.0333 	&	0.1655 	&	0.4395 	&	0.1655 	&	0.4407 	&	0.0013 	&	-0.0025 	&	0.0134 	&	0.0275 	&	0.0135 	&	0.0276 	\\
	$\delta_{10}$	&	0.0129 	&	0.0227 	&	0.2394 	&	0.4157 	&	0.2398 	&	0.4164 	&	-0.0014 	&	-0.0048 	&	0.0137 	&	0.0309 	&	0.0138 	&	0.0312 	\\
	$\delta_{11}$	&	0.0282 	&	0.0244 	&	0.1698 	&	0.3892 	&	0.1721 	&	0.3900 	&	-0.0007 	&	-0.0036 	&	0.0172 	&	0.0307 	&	0.0172 	&	0.0309 	\\
	$\delta_{12}$	&	0.0104 	&	0.0096 	&	0.1320 	&	0.3395 	&	0.1324 	&	0.3396 	&	-0.0012 	&	-0.0034 	&	0.0092 	&	0.0264 	&	0.0093 	&	0.0267 	\\
	$\delta_{13}$	&	0.0091 	&	0.0355 	&	0.1261 	&	0.3502 	&	0.1264 	&	0.3520 	&	-0.0001 	&	0.0009 	&	0.0154 	&	0.0313 	&	0.0154 	&	0.0314 	\\
	$\delta_{14}$	&	0.0024 	&	0.0241 	&	0.1866 	&	0.4005 	&	0.1866 	&	0.4013 	&	0.0005 	&	0.0026 	&	0.0149 	&	0.0283 	&	0.0149 	&	0.0284 	\\
	$\delta_{15}$	&	0.0072 	&	0.0219 	&	0.2171 	&	0.4147 	&	0.2172 	&	0.4153 	&	0.0011 	&	0.0024 	&	0.0140 	&	0.0290 	&	0.0140 	&	0.0291 	\\	
	\midrule
	\multicolumn{7}{l}{Panel C1: $T=400$}& \multicolumn{6}{|l}{Panel C2: $T=400$} \\
	$\alpha_0$	&	0.1549 	&	0.2920 	&	0.3905 	&	0.7234 	&	0.4201 	&	0.7801 	&	0.0015 	&	-0.0017 	&	0.0284 	&	0.0456 	&	0.0285 	&	0.0456 	\\
	$\beta_0$	&	-0.1174 	&	-0.1514 	&	0.5202 	&	1.1991 	&	0.5333 	&	1.2087 	&	0.0046 	&	0.0100 	&	0.0740 	&	0.1204 	&	0.0741 	&	0.1208 	\\
	$\delta_1$	&	0.0074 	&	0.0079 	&	0.2311 	&	0.2536 	&	0.2312 	&	0.2537 	&	0.0000 	&	-0.0001 	&	0.0180 	&	0.0189 	&	0.0180 	&	0.0189 	\\
	$\delta_2$	&	0.0404 	&	0.0425 	&	0.2231 	&	0.2487 	&	0.2268 	&	0.2523 	&	-0.0040 	&	-0.0044 	&	0.0206 	&	0.0213 	&	0.0210 	&	0.0218 	\\
	$\delta_3$	&	0.0450 	&	0.0457 	&	0.2226 	&	0.2352 	&	0.2271 	&	0.2396 	&	0.0001 	&	-0.0002 	&	0.0189 	&	0.0197 	&	0.0189 	&	0.0197 	\\
	$\delta_4$	&	0.0354 	&	0.0335 	&	0.2177 	&	0.2271 	&	0.2206 	&	0.2295 	&	0.0030 	&	0.0023 	&	0.0182 	&	0.0190 	&	0.0185 	&	0.0192 	\\
	$\delta_5$	&	-0.0109 	&	-0.0069 	&	0.2604 	&	0.2765 	&	0.2607 	&	0.2765 	&	-0.0002 	&	0.0017 	&	0.0199 	&	0.0203 	&	0.0199 	&	0.0204 	\\
	$\delta_6$	&	-0.0042 	&	0.0090 	&	0.1346 	&	0.2682 	&	0.1346 	&	0.2684 	&	-0.0003 	&	0.0010 	&	0.0109 	&	0.0172 	&	0.0109 	&	0.0172 	\\
	$\delta_7$	&	-0.0072 	&	0.0129 	&	0.1265 	&	0.2419 	&	0.1267 	&	0.2423 	&	-0.0014 	&	-0.0015 	&	0.0138 	&	0.0195 	&	0.0139 	&	0.0195 	\\
	$\delta_8$	&	0.0174 	&	0.0045 	&	0.1745 	&	0.2883 	&	0.1754 	&	0.2883 	&	0.0019 	&	0.0025 	&	0.0134 	&	0.0215 	&	0.0135 	&	0.0216 	\\
	$\delta_9$	&	0.0047 	&	0.0059 	&	0.1469 	&	0.2530 	&	0.1469 	&	0.2530 	&	0.0000 	&	0.0009 	&	0.0092 	&	0.0163 	&	0.0092 	&	0.0163 	\\
	$\delta_{10}$	&	0.0147 	&	0.0223 	&	0.1296 	&	0.2568 	&	0.1304 	&	0.2578 	&	-0.0001 	&	-0.0005 	&	0.0135 	&	0.0198 	&	0.0135 	&	0.0198 	\\
	$\delta_{11}$	&	-0.0020 	&	0.0118 	&	0.1774 	&	0.2897 	&	0.1774 	&	0.2899 	&	-0.0024 	&	-0.0038 	&	0.0137 	&	0.0186 	&	0.0139 	&	0.0190 	\\
	$\delta_{12}$	&	-0.0154 	&	-0.0244 	&	0.1291 	&	0.2211 	&	0.1300 	&	0.2224 	&	0.0002 	&	-0.0009 	&	0.0141 	&	0.0206 	&	0.0141 	&	0.0206 	\\
	$\delta_{13}$	&	0.0044 	&	-0.0027 	&	0.1195 	&	0.2298 	&	0.1196 	&	0.2298 	&	-0.0005 	&	0.0011 	&	0.0125 	&	0.0180 	&	0.0125 	&	0.0180 	\\
	$\delta_{14}$	&	0.0079 	&	0.0005 	&	0.1222 	&	0.2324 	&	0.1225 	&	0.2324 	&	-0.0018 	&	-0.0018 	&	0.0144 	&	0.0209 	&	0.0145 	&	0.0209 	\\
	$\delta_{15}$	&	0.0035 	&	-0.0103 	&	0.1092 	&	0.2352 	&	0.1093 	&	0.2354 	&	-0.0022 	&	-0.0013 	&	0.0109 	&	0.0183 	&	0.0111 	&	0.0184 	\\
	$\delta_{16}$	&	-0.0122 	&	-0.0108 	&	0.1712 	&	0.2848 	&	0.1716 	&	0.2850 	&	-0.0006 	&	-0.0013 	&	0.0144 	&	0.0237 	&	0.0145 	&	0.0237 	\\
	$\delta_{17}$	&	0.0142 	&	0.0038 	&	0.1322 	&	0.2658 	&	0.1330 	&	0.2658 	&	-0.0021 	&	0.0000 	&	0.0131 	&	0.0210 	&	0.0132 	&	0.0210 	\\
	$\delta_{18}$	&	0.0187 	&	0.0106 	&	0.1378 	&	0.2854 	&	0.1391 	&	0.2856 	&	0.0008 	&	0.0033 	&	0.0121 	&	0.0187 	&	0.0121 	&	0.0190 	\\
	$\delta_{19}$	&	0.0127 	&	0.0152 	&	0.1348 	&	0.2627 	&	0.1354 	&	0.2632 	&	0.0008 	&	-0.0011 	&	0.0130 	&	0.0214 	&	0.0130 	&	0.0215 	\\
	$\delta_{20}$	&	-0.0067 	&	-0.0310 	&	0.1036 	&	0.2163 	&	0.1038 	&	0.2185 	&	0.0004 	&	0.0007 	&	0.0128 	&	0.0193 	&	0.0128 	&	0.0193 	\\	
	\bottomrule
	\end{tabular}
	\caption*{\footnotesize Note: Our method is denoted as PLR, and the minimum $D_k$-distance estimation of ACIX is denoted as ACIX. In Panel A1, B1, and C1 of this table, the interval innovation error is generated by ACI model. Additionally, all the values in this panel are obtained by multiplying the original value by $10^3$. In Panel A2, B2, and C2, the interval innovation error is generated by bivariate normal distribution.  In addition, $(a,b,c)=(5,1,1)$ and $\gamma=1$.}
\label{tab_DGP2}
\end{table}




\section{Empirical applications}
\subsection{Interval-valued crude oil price forecasting}
Accurate crude oil price forecasting is an important yet controversial issue in economic and management research. Numerous studies have demonstrated that crude oil prices are influenced by a myriad of financial and macroeconomic factors, such as supply and demand dynamics, stock market performance, interest rates, exchange rates, monetary policy, and other commodity prices \citep{NASER201675, WEI2017141, he2021forecasting, he2010empirical}. Importantly, many of these factors are represented in the form of interval-valued data, reflecting the inherent uncertainty and variability in their measurement. However, most existing work has focused on point-valued data, potentially overlooking valuable information contained within the interval-valued representations. Modeling interval-valued data may capture more comprehensive information, thereby enhancing the accuracy of crude oil price predictions. Consequently, we are motivated to identify and select important interval-valued factors from various potential variables via shrinkage methods, with the expectation of improving the forecasting accuracy of interval-valued crude oil prices.

\subsubsection{Data description}
\begin{figure}[h!]
	\centering
	\vskip -0.15in
	\includegraphics[width=0.6\textwidth]{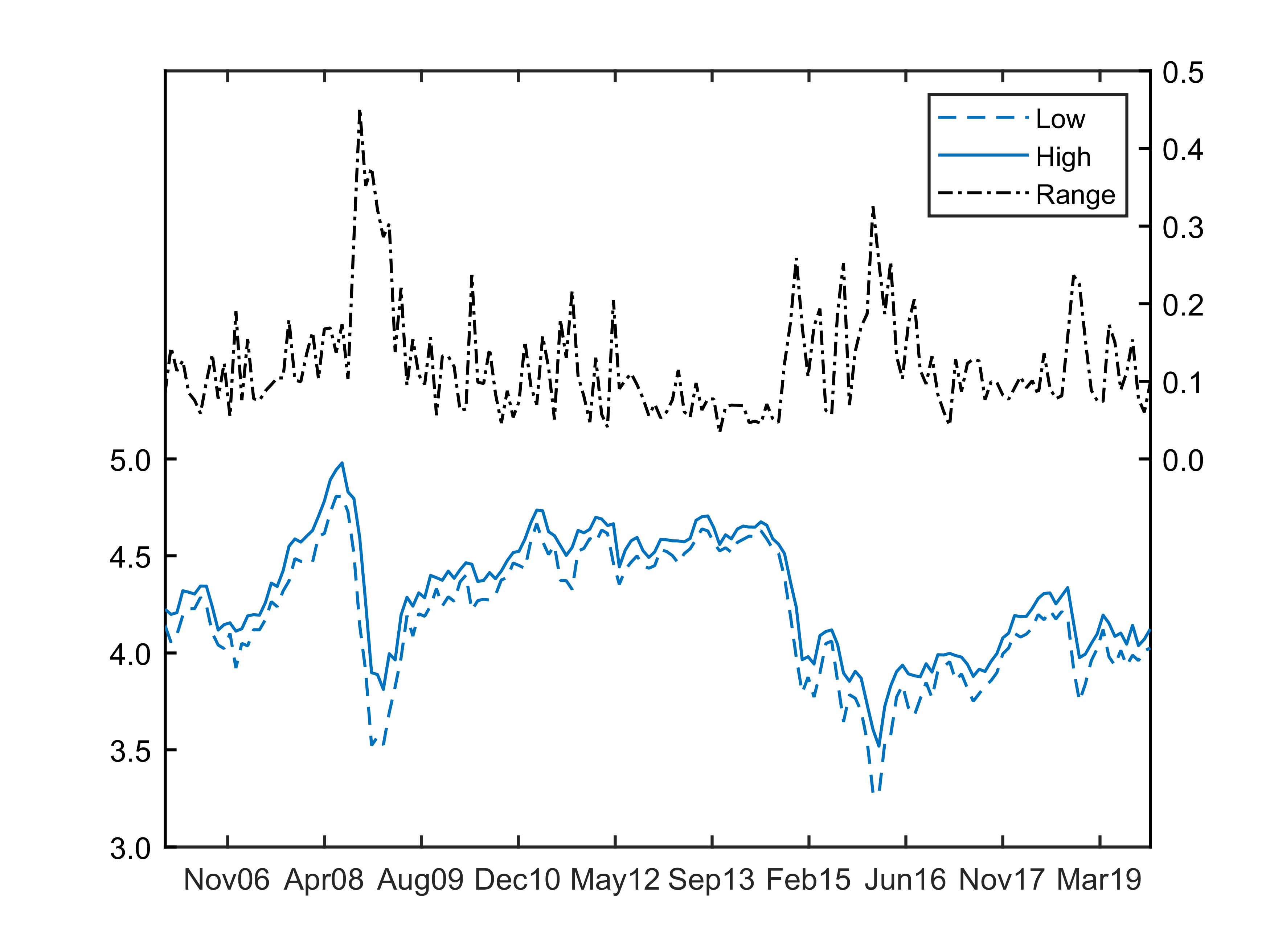}
	\vskip -0.25in
	\caption{\footnotesize The monthly interval-valued WTI crude oil futures prices. The sample period is from January 2006 to December 2019. The dotted line, solid line and chain line respectively represent ``Low" (logarithmic low price, $Y_{L,t}$), ``High" (logarithmic high price, $Y_{R,t}$) and ``Range" ($Y_{R,t}-Y_{L,t}$). The left vertical axis is for ``Low" and ``High", and the right one is for ``Range".}
	\label{fig1}
\end{figure}

This section describes the data used in our analysis, focusing on the monthly interval-valued West Texas Intermediate (WTI) crude oil futures prices. These prices are constructed from daily closing prices sourced from the New York Mercantile Exchange (NYMEX), a major marketplace for crude oil futures trading. Denote the ITS of crude oil prices as $Y_t=[Y_{L,t},Y_{R,t}]$, where $Y_{L,t}$ and $Y_{R,t}$ are constructed by taking the logarithm of the minimum and maximum daily closing prices within month $t$, respectively. The sample period spans from January 2006 to December 2019. Figure \ref{fig1} illustrates the bounds and range of these interval-valued prices over time.
\vspace{-10pt}

We draw some interesting observations from Figure \ref{fig1}. First, the ITS of crude oil price captures intra-month variations that monthly closing prices fail to reflect. Additionally, there appears to be a strong correlation between the lower and upper bounds of the interval. Second, the range of oil prices tends to increase as the price level decreases, suggesting that crude oil prices may become more volatile during downward trends. This pattern demonstrates that price level and volatility are two distinct aspects of crude oil price movements, likely exhibiting a negative correlation. Third, two considerable drops are evident in the trend of interval-valued prices. The first occurs from July to December 2008, attributable to the subprime crisis, while the second spans from June 2014 to January 2016, resulting from shale oil shocks.

As a crucial strategic resource on a global scale, crude oil typically exhibits high volatility in its pricing, influenced by a myriad of factors. Table \ref{tab:factors} presents several factors used in our application, including stock market indicators, monetary market variables, and supply and demand metrics. These factors have been widely used in existing literature \citep{wang2016forecasting,chai2018forecasting,Yang2016}. To assess the stationarity of our data, we applied the Augmented Dickey-Fuller (ADF) test to $Y_{L,t}$, $Y_{R,t}$, and the bounds of explanatory variables. The results indicate that all point-valued series are first-order stationary.
\begin{table}[htbp]
\scriptsize
	\centering
	\caption{Explanatory variables}
	\begin{tabular}{l|l}
	  \toprule
	  Stock market & Dow Jones Industrial Average (DJIA) \\
	  \midrule
	  Commodity market & log copper future price in London metal exchange (LME) \\
	  \midrule
	  Monetary market
	   & generalized nominal dollar index \\
	   & long-term average real interest rate of U.S. treasury bonds \\
	   & U.S. interest rate on a three-month Treasury bill (secondary market) \\
	   & U.S. M2 \\
	  \midrule
	  Crude oil supply and demand
	   & log U.S. imports of crude oil \\
	   & log U.S. production of crude oil \\
	   & log U.S. ending stocks of crude oil \\
	  \midrule
	  Technology & WTI-Brent future closing price spread \\
	  \midrule
	  Speculation & interval-valued speculation index (see details in \cite{Yang2016}) \\
	  \bottomrule
	\end{tabular}
	\label{tab:factors}
\end{table}

\subsubsection{Evaluation} \label{sec EVA}
To explore the performance of sparse model for ITS, we first employ the ACIX model \citep{han2016vector} with all predictors as a benchmark forecasting model. Next, following the spirit of \cite{Gloria2013Constrained} and \cite{SUN2021MA}, the center-range method (CRM) and the constrained center-range method (CCRM) are considered as benchmark forecast methods. These two methods first proposed by \cite{LIMANETO20081500,LIMANETO2010333} can be expressed as: $y_t^m = \beta_0^m+\beta_1^m x_{1,t}^m + \cdots +\beta_p^m x_{p,t}^m + \epsilon_t^m$ and $y_t^r = \beta_0^r+\beta_1^r x_{1,t}^r + \cdots +\beta_p^r x_{p,t}^r + \epsilon_t^r$,
where $\{y_t^m, x_{i,t}^m\}$ are midpoints of interval observations, and $\{y_t^r, x_{i,t}^r\}$ represent the ranges. Both CRM and CCRM estimators can be obtained from two separate point-valued least squares estimation methods. CCRM imposes restrictions on the coefficients of range, i.e., $\beta_i^r\geq 0$ for $1\leq i\leq p$, to ensure the range variables are nonnegative. A bivariate model for lower and upper bounds (BLU) is also employed as a benchmark forecast model. BLU model is based on estimation in the following system: $y_t^l = \beta_0^l+\beta_1^l x_{1,t}^l + \cdots +\beta_p^l x_{p,t}^l + \epsilon_t^l$ and $y_t^u = \beta_o^u+\beta_1^u x_{1,t}^u + \cdots +\beta_p^u x_{p,t}^u + \epsilon_t^u$.
{Furthermore, to evaluate the performance of our proposed interval-based adaptive LASSO model relative to alternative machine learning approaches, we utilize random forest and MLP models as benchmarks for interval-valued data, denoted as IRF and IMLP\footnote{The random forest model is constructed with 100 decision trees using the TreeBagger algorithm. The MLP uses a single hidden layer with 10 neurons, trained for maximum 1000 epochs with error goal of $10^{-5}$ and minimum gradient of $10^{-6}$, with a 70:15:15 data split ratio for training, validation and testing.}. They employ a strategy of estimating the lower and upper bounds of intervals separately.}

The evaluation criteria listed in Table \ref{tab criteria} have frequently been used in interval regression models \citep{maciel2017evolving, Rodrigues2015, maciel2023adaptive, yang2024LSTIAR} to evaluate the forecast accuracy of these models. The criteria in Panel A measure the gap between predicted and actual intervals, while the other criteria in Panel B measure forecast accuracy of special points in predicted intervals. Specifically, $\omega_1$ measures the nonoverlapping area of the forecasting and actual intervals, and $\omega_{D_K}$ measures the $D_K$-distance between $\hat{Y}_t$ and $Y_t$. $\omega_{NSD1}$ and $\omega_{NSD2}$ evaluate the nonoverlapping area of $\hat{Y}_t$ and $Y_t$ relative to their union set, where $\omega(\cdot)$ denotes the width and $R(\cdot)$ the range of an interval, see more details in \cite{sun2018threshold}. Moreover, $\omega_{MDE}$ is about the mean distance error. In addition, $\omega_{rate}$ refers to the non-efficiency rate. For the point-based criteria, all four are the special cases of RMSE.
\begin{table}[htbp]
	\scriptsize
	\centering
	\caption{The evaluation criteria}
	\begin{tabular}{l|l}
	\toprule
	\emph{Panel A: }Interval-based criteria & Meaning \\
	\midrule
	\rule{0pt}{15pt}
	$\omega_1 = 1-\frac{1}{T}\sum_t\frac{\min(\hat{H}_t,H_t)-\max(\hat{L}_t,L_t)}{\max(\hat{H}_t,H_t)-\min(\hat{L}_t,L_t)}$ & nonoverlapping area \\
	\midrule
	\rule{0pt}{15pt}
	$\omega_{D_K} = \frac{1}{T}\sqrt{\sum_t D_K(\hat{Y}_t,Y_t)}$ & forecast accuracy based on $D_K$-distance \\
	\midrule
	\rule{0pt}{15pt}
	$\omega_{NSD1} = \frac{1}{T}\sum_t\frac{w([L_t,H_t]\cup[\hat{L}_t,\hat{H}_t])-w([L_t,H_t]\cap[\hat{L}_t,\hat{H}_t])}{w([L_t,H_t]\cup[\hat{L}_t,\hat{H}_t])}$ & \multirow{2}{*}{normalized symmetric difference of intervals}\\
	\rule{0pt}{15pt}
	$\omega_{NSD2} = 2-\frac{1}{T}\sum_t\frac{w([\hat{L}_t,\hat{H}_t])+w([L_t,H_t])}{R([L_t,H_t]\cup[\hat{L}_t,\hat{H}_t])}$ & \\
	\midrule
	\rule{0pt}{15pt}
	$\omega_{MDE} = \frac{1}{T}\sum_t\sqrt{|\hat{M}_t-M_t|^2+|\hat{R}_t-R_t|}$ & mean distance error\\
	\midrule
	\rule{0pt}{15pt}
	$\omega_{rate}=1-\frac{1}{T}\sum_t\frac{w([L_t,H_t]\cap[\hat{L}_t,\hat{H}_t])}{w([\hat{L}_t,\hat{H}_t])}$ & non-efficiency rate \\
	\midrule
	\emph{Panel B: }Point-based criteria & \\
	\midrule
	\rule{0pt}{15pt}
	$\omega_M = \sqrt{\frac{1}{T}\sum_t(\hat{M}_t-M_t)^2}$ & RMSE of midpoint \\
	\rule{0pt}{15pt}
	$\omega_R = \sqrt{\frac{1}{T}\sum_t(\hat{R}_t-R_t)^2}$ & RMSE of radius \\
	\rule{0pt}{15pt}
    $\omega_L = \sqrt{\frac{1}{T}\sum_t(\hat{L}_t-L_t)^2}$ & RMSE of lower bound \\
	\rule{0pt}{15pt}
	$\omega_H = \sqrt{\frac{1}{T}\sum_t(\hat{H}_t-H_t)^2}$ & RMSE of upper bound \\
	\bottomrule
	\end{tabular}
	\caption*{\footnotesize Note: $\hat{Y}_t=[\hat{L}_t,\hat{H}_t]$ is the predicted interval of crude oil price in month $t$, and $Y_t=[L_t,H_t]$ is the actual interval. $M_t$ $(\hat{M}_t)$ and $R_t$ $(\hat{R}_t)$ are the midpoint and radius of the actual (predicted) intervals.}
	\label{tab criteria}
\end{table}
\vspace{-0.5cm}
\subsubsection{Forecasting performance}
We employ a rolling window approach to study the out-of-sample performance of our estimation and other interval-based methods. A rolling estimation scheme is adopted for $60$ months with the first estimation sample spanning from January 2006 to December 2010. We conduct the one-step-ahead out-of-sample forecasts from January 2011 to December 2019, including $T_f=108$ forecasting periods. In addition, we also adopt a rolling estimation scheme for $120$ months, namely from January 2006 to December 2015. The last observations are forecasting sample. For each fixed rolling window, we compare our method's performance with that of the benchmark methods.

Table \ref{tab:oil-interval} shows the out-of-sample performance of seven interval forecasting methods: ACIX, CRM, CCRM, BLU, IRF, IMLP, and our proposed method, focusing on interval-based criteria as outlined in Panel A of Table \ref{tab criteria}. Among the seven interval-based methods considered, our method consistently ranks highest in forecasting performance across all interval-based criteria. For instance, with the training sample size of 60, our model's $D_K$-distance measure $w_{D_K}$ is 0.0086, outperforming all the other values. This superior performance can be attributed to two main factors. First, our interval model treats the interval oil price sample as an inseparable set and employs information of distances between both boundaries and interior points. Particularly, the correlation between the interval's lower and upper bounds is taken into account. As Figure \ref{fig1} shows, because the bounds of interval crude oil price are not independent, modeling the interval's lower and upper bounds separately does not fully utilize the information of the intervals. Second, the penalized interval regression provides a sparse model with fewer explanatory variables, effectively excluding the influence of unrelated and weakly related variables. Compared with our interval-based machine learning method, the underperformance of classic machine learning algorithms, i.e., IRF and IMLP, in predicting interval-valued oil prices can be attributed to two main factors. First, they simply apply regression to the two endpoints of the interval, neglecting the internal information. Second, the relatively small sample size could lead to overfitting in these complex models, particularly if not adequately tuned for the specific dataset.
\begin{table}[ht!]
    \scriptsize
    \centering
    \caption{Out-of-sample forecast evaluation: Interval-based criteria}
    \begin{tabular}{lcccccc}
        \toprule
        & $\omega_1$ & $\omega_{D_K}$ & $\omega_{NSD1}$ & $\omega_{NSD2}$ & $\omega_{MDE}$ & $\omega_{rate}$ \\
        \midrule
        \textbf{\emph{Panel A}} \\
        PLR & \textbf{0.4629} & \textbf{0.0086} & \textbf{0.4592} & \textbf{0.4592} & \textbf{0.0372} & \textbf{0.3153} \\
        ACIX & 0.5592 & 0.0108 & 0.5425 & 0.5425 & 0.0468 & 0.4052 \\
        CRM & 0.5469 & 0.0141 & 0.5370 & 0.5370 & 0.0595 & 0.4986 \\
        CCRM & 0.5399 & 0.0138 & 0.5310 & 0.5310 & 0.0582 & 0.4929 \\
        BLU & 0.6222 & 0.0123 & 0.5649 & 0.5649 & 0.0533 & 0.4044 \\
        IRF & 0.7668 & 0.0218 & 0.7166 & 0.7166 & 0.0862 & 0.6037 \\
        IMLP & 1.1366 & 0.0331 & 0.7300 & 0.7300 & 0.1136 & 0.5978 \\
        \midrule
        \textbf{\emph{Panel B}} \\
        PLR & \textbf{0.3818} & \textbf{0.0110} & \textbf{0.3818} & \textbf{0.3818} & \textbf{0.0305} & \textbf{0.2399} \\
        ACIX & 0.5169 & 0.0155 & 0.5034 & 0.5034 & 0.0447 & 0.3561 \\
        CRM & 0.4678 & 0.0206 & 0.4674 & 0.4674 & 0.0504 & 0.4361 \\
        CCRM & 0.4664 & 0.0201 & 0.4664 & 0.4664 & 0.0497 & 0.4334 \\
        BLU & 0.4788 & 0.0142 & 0.4759 & 0.4759 & 0.0425 & 0.3399 \\
        IRF & 0.7341 & 0.0303 & 0.6930 & 0.6930 & 0.0853 & 0.5746 \\
        IMLP & 0.8881 & 0.0432 & 0.6661 & 0.6661 & 0.0954 & 0.5159 \\
        \bottomrule
    \end{tabular}
    \vspace{0.5em}
    \caption*{\footnotesize Note: ``PLR" means our proposed model, i.e., the penalized interval regression model. All criteria are between 0 and 1; the smaller, the better. The smallest value in each column for each panel is highlighted in bold. In \emph{Panel A}, the sample size of training data for rolling window estimation is 60. In \emph{Panel B}, the sample size of training data for rolling window estimation is 120. The initial parameters are set as $(a,b,c)=(5,1,1)$ and $\gamma=0.5$.}
    \label{tab:oil-interval}
\end{table}

The point-based criteria are outlined in Table \ref{tab oil point}. It is observed that our method consistently outperforms the other six benchmark interval methods across all point-based criteria. To verify these findings, a Diebold-Mariano (DM) test is conducted on the out-of-sample forecast performance of various points within the forecasting intervals (including lower and upper boundaries, midpoint, and radius). The results of the DM tests, denoted by asterisks in Table \ref{tab oil point}, provide strong evidence of the superiority of our PLR method. For example, in Panel A, we observe that our sparse method significantly outperforms all benchmark models at the 1\% level for nearly all criteria. The only exception is the radius measure for the IRF model, where the difference is significant at the 5\% level. Panel B results remain significant, albeit less so than Panel A. This difference likely stems from a smaller test set sample in Panel B, reducing statistical power.
\begin{table}[ht!] 
    \scriptsize
    \centering
    \caption{Out-of-sample forecast evaluation: Point-based criteria and DM test}
    \begin{tabular}{lllll}
    \toprule
    $\theta$ & Lower bound & Upper bound & Midpoint & Radius \\
    \midrule
    \textbf{\emph{Panel A}}\\
     PLR    &0.0517      &0.0405      &0.0426     &0.0185 \\
     ACIX   &$0.0641^{***}$ &$0.0512^{***}$ &$0.0536^{***}$ &$0.0219^{***}$ \\
     CRM    &$0.0752^{***}$ &$0.0651^{***}$ &$0.0628^{***}$ &$0.0317^{***}$ \\
     CCRM   &$0.0734^{***}$ &$0.0647^{***}$ &$0.0628^{***}$ &$0.0291^{***}$ \\
     BLU    &$0.0774^{***}$ &$0.0536^{***}$ &$0.0573^{***}$ &$0.0305^{***}$ \\
     IRF    &$0.1284^{***}$ &$0.1095^{***}$ &$0.1169^{***}$ &$0.0232^{**}$ \\
     IMLP   &$0.1701^{***}$ &$0.1317^{***}$ &$0.1073^{***}$ &$0.0896^{***}$ \\
    \midrule
    \textbf{\emph{Panel B}}\\
    PLR     &0.0402      &0.0338      &0.0338     &0.0154 \\
    ACIX    &$0.0599^{***}$ &$0.0482^{***}$ &$0.0501^{***}$ &$0.0208^{***}$ \\
    CRM     &$0.0497^{**}$ &$0.0640^{***}$ &$0.0500^{***}$ &$0.0279^{***}$ \\
    CCRM    &$0.0495^{*}$ &$0.0630^{***}$ &$0.0500^{***}$ &$0.0265^{***}$ \\
    BLU     &$0.0611^{***}$ &$0.0431^{***}$ &$0.0483^{***}$ &$0.0215^{**}$ \\
    IRF     &$0.1171^{***}$ &$0.0998^{***}$ &$0.1059^{***}$ &$0.0250^{**}$ \\
    IMLP    &$0.1393^{***}$ &$0.1186^{***}$ &$0.0966^{***}$ &$0.0737^{**}$ \\
    \bottomrule
    \end{tabular}
    \caption*{\footnotesize Note: \emph{Panel A, B} and the initial parameters are set the same as Table \ref{tab:oil-interval}. Four point-based criteria $w_L$, $w_H$, $w_M$ , and $w_R$ are reported in this table, corresponding to the RMSE of lower/upper bound, midpoint, and radius. The superscript $^{*}$, $^{**}$ and $^{***}$ represent significance levels of 0.1, 0.05 and 0.01 in the DM test, respectively.}
    \label{tab oil point}
\end{table}

\subsection{Interval-based index tracking}
Index tracking is a crucial technique in portfolio management, enabling investors to replicate the performance of a benchmark index while minimizing tracking error. Traditionally, most existing literature on index tracking have primarily relied on closing prices as the input data source \citep{corielli2006factor,wu2014nonnegative,strub2018optimal}. However, these approaches may fail to capture the full extent of price fluctuations that occur throughout the trading day. Neglecting intraday price movements can potentially lead to suboptimal portfolio construction and increased tracking error, as the closing price alone may not accurately reflect the true dynamics of the underlying assets, which is a critical aspect in index tracking. To address this issue, incorporating interval-valued data can provide a more comprehensive representation of asset price movements, which can better account for volatility and price variations, potentially leading to improved tracking performance and reduced tracking error. Consequently, we are motivated to develop interval-based index tracking methodologies, aiming to construct portfolios that more accurately replicate the benchmark index's performance. To the best of our knowledge, this paper is the first to propose a replicating strategy based on interval-valued stock prices, marking a novel contribution to the field of index tracking.

\subsubsection{Strategies of index tracking}
\begin{figure}[htbp]
	\centering
	\includegraphics[width=0.5\textwidth]{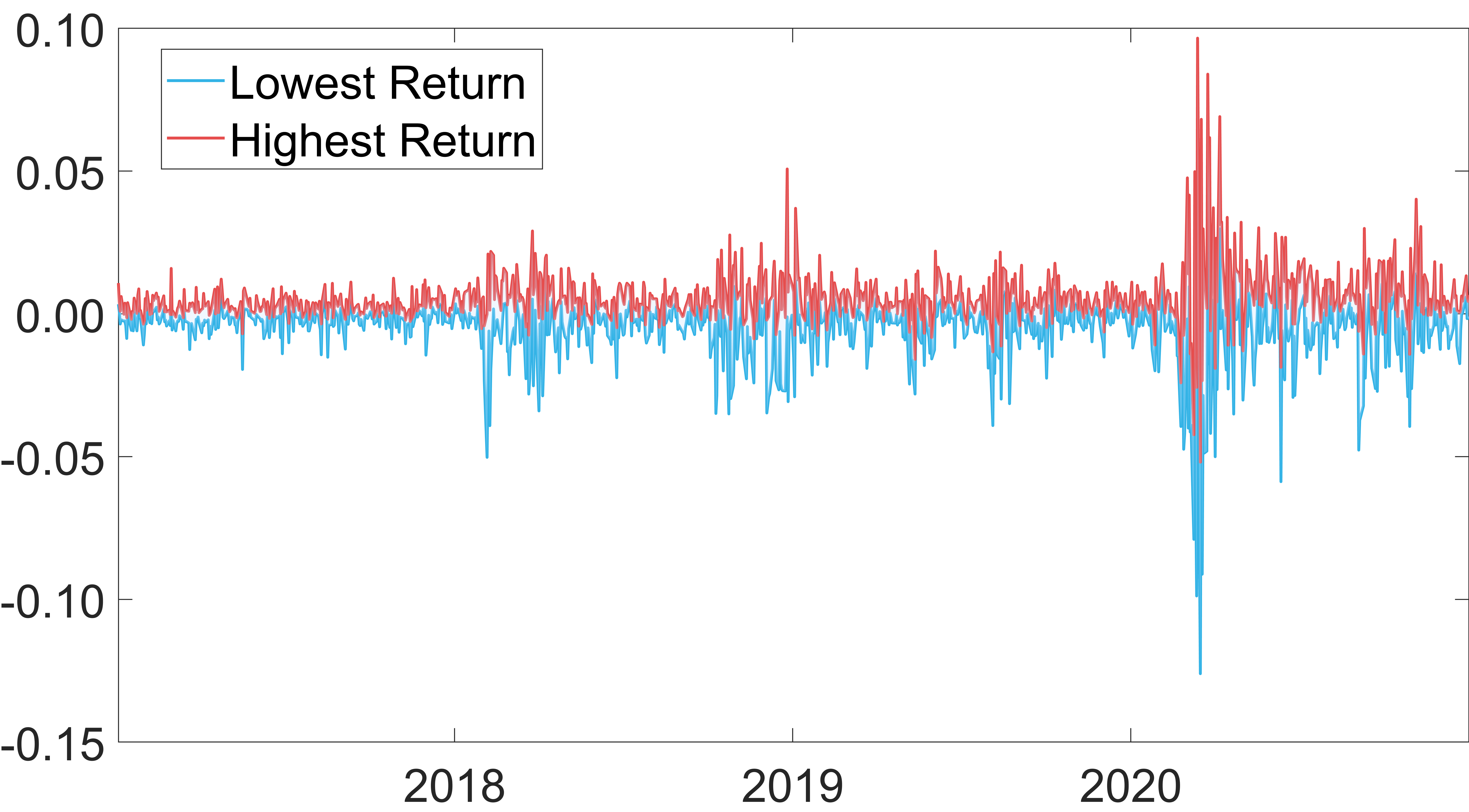}
	\caption{\footnotesize The daily interval-valued log returns of the S\&P 100 index. The sample period is from 03/01/2017 to 31/12/2020. The blue and the red line represent the low and the high log returns of the S\&P 100 index, respectively.}
	\label{fig2}
\end{figure}

In this application, we develop and evaluate an interval-based strategy for tracking the S\&P 100 index. The interval-valued log return of stocks is constructed as $[r_{l,t},r_{h,t}]$, where $r_{l,t}=\ln\frac{P_{low,t}}{P_{close,t-1}}$ and $r_{h,t}=\ln\frac{P_{high,t}}{P_{close,t-1}}$. This representation of daily interval-valued returns, also adopted by \cite{Gloria2013Constrained}, captures investors' high and low expectations and provides more comprehensive trading information. As shown in Figure \ref{fig2}, the S\&P 100 index returns demonstrated notable volatility, particularly during the COVID-19 pandemic period.

Our index tracking strategy comprises two main steps. First, we employ the proposed penalized linear interval regression to select stocks from the S\&P 100 constituents, with the number of selected stocks controlled by the tuning parameter. Second, we estimate the weights of the chosen stocks using OLS regression on closing prices.\footnote{Our strategy does not enforce full investment (sum of weights needs not equal one) and allows short selling, enabling direct OLS estimation of weights. See \cite{shu2020high} for details about this approach.} For comparison, we construct a point-based benchmark strategy that follows the same two-step procedure but uses LASSO \citep{tibshirani1996regression} for stock selection, with closing prices utilized in both steps.

To implement and evaluate these strategies, we use a rolling window approach with a 250-day training period (approximately one trading year) and a 21-day testing period (approximately one trading month). The process involves selecting 10 stocks from the S\&P 100 constituents using both interval-based and point-based methods, followed by weight estimation through OLS regression. To ensure robustness, we examine three different training samples beginning from the first trading day of 2017, 2018, and 2019, respectively, comparing both in-sample and out-of-sample performance. All data are sourced from Wind and Yahoo Finance.

To evaluate the performance of our interval-based strategy and the point-based strategy, we employ the following two widely used measures of tracking error \citep{wu2014nonnegative,corielli2006factor}:
\begin{align*}
    &S(T) = \sqrt{\frac{1}{T-1}\sum_{t=1}^T(err_t-\overline{err})^2}, \\
    &M(T) = \frac{1}{T}\sum_{t=1}^T |err_t|,
\end{align*}
where $T$ is the sample size, $err_t = r_t - \hat{r}_t$, $\overline{err}$ is the sample mean of $err_t$, and $r_t$ and $\hat{r}_t$ are the returns from the tracking portfolio and the index, respectively.

\subsubsection{Tracking performance}
\begin{figure}[h!]
	\centering
	\includegraphics[width=0.8\textwidth]{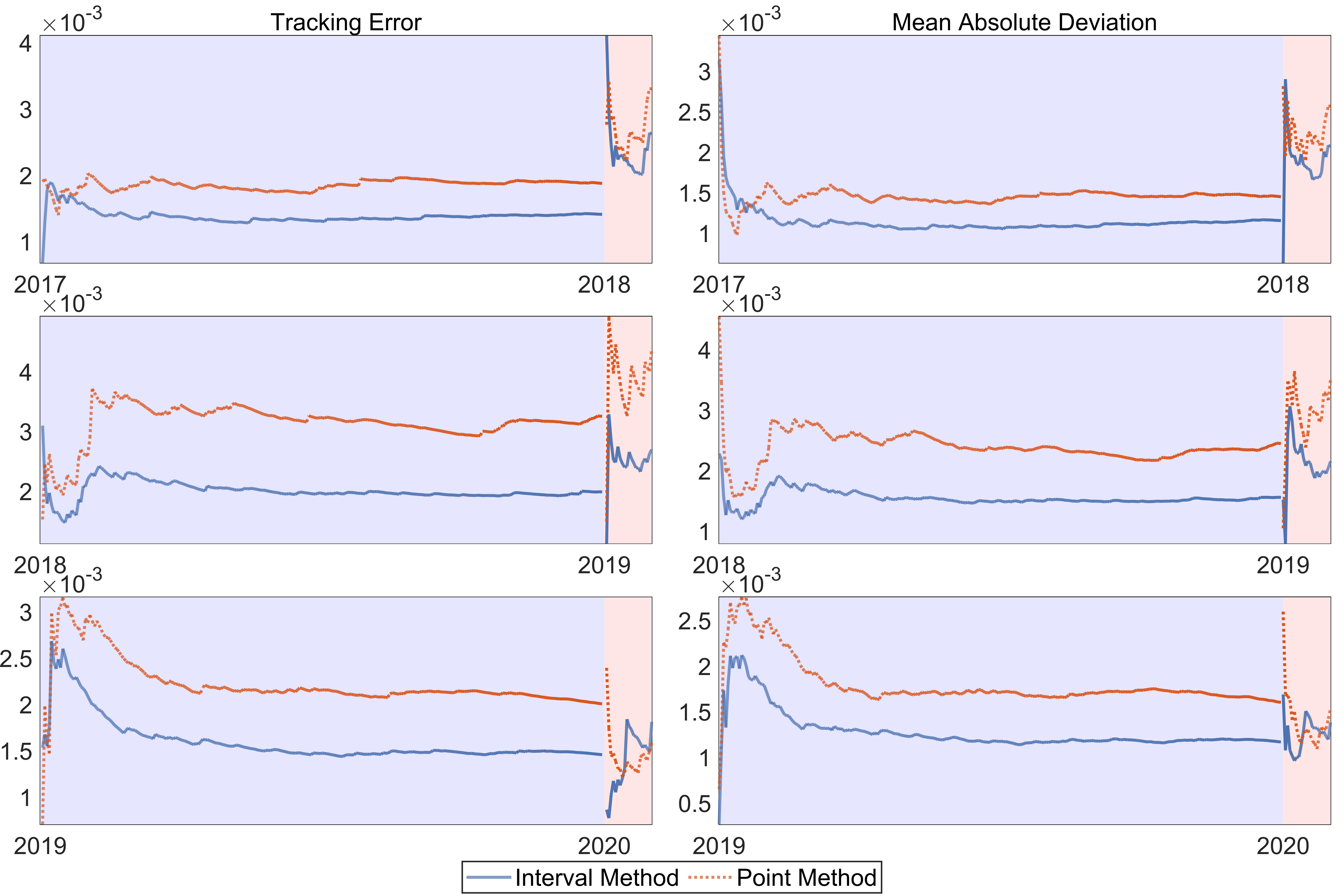}
	\caption{\footnotesize The blue solid lines represent the interval-based index tracking strategy, while the red dashed lines refer to the point-based methods. The background colors indicate different sample periods: light blue represents the in-sample period, while light red represents the out-of-sample period. For parameter setting, the initial kernel is set to be $(5,1,1)$, and $\gamma=0.5$.}
	\label{fig3}
\end{figure}

Figure \ref{fig3} displays the tracking error and mean absolute deviation of both the interval-based and point-based models. To offer more detailed insights, we illustrate the cumulative error in Figure \ref{fig3}, where the lines represent $S(\tau)$ and $M(\tau)$ as $\tau$ ranges from 1 to $250$ in the training sample and from 1 to 21 in the test sample.\footnote{
To avoid confusion, it is necessary to clarify that the model is estimated from the entire set of 250 training samples, and we only vary $\tau$ during evaluation.}

Several observations can be obtained from Figure \ref{fig3}. First, the interval-based strategy outperforms the point-based one in-sample, as measured by both types of tracking errors. At the beginning, the blue line occasionally exceeds the red line, which is attributed to the instability resulting from the limited sample size. As the sample size used for calculating cumulative error gradually increases, the in-sample performance of our method notably outperforms that of the point-based method. Second, the out-of-sample tracking errors in 2018 and 2019 of the interval-based method are mostly smaller than those of the point-based method. The outperformance of our proposed interval-based index tracking strategy illustrates that interval-valued data may contribute to improving conventional portfolio strategies. Third, when the out-of-sample is 2020, neither method demonstrates a significant advantage over the other. One possible reason is the extreme volatility in the stock market caused by the COVID-19 pandemic in early 2020, making it challenging for past performance to capture market behavior during such extreme shocks.
Overall, the application of the proposed interval variable selection methods in index tracking demonstrates that interval-valued data can improve portfolio strategies. This inspires us to develop deeper research of interval models in financial studies in the future work.

\section{Conclusion}
In this paper, we propose a sparse regression  for high-dimensional interval-valued data via machine learning tools.
It is shown that the proposed method enjoys the oracle properties, i.e., the consistency in variable selection and the asymptotic normality of estimators. We further extend the proposed method to a diverging-dimensional interval case. Additionally, we also propose an interval-based LARS algorithm to solve the solution path of the estimation. Furthermore, simulation studies confirm the asymptotic properties of our method. Empirical applications highlight that our method improves the out-of-sample forecasts of crude oil price and works well in index tracking.

The proposed machine learning technique for interval linear regression is an interval-valued extension of the adaptive LASSO, originally designed for point-valued data. Several important avenues for future research emerge from this work. One potential extension is to adopt more sophisticated and powerful machine learning techniques, such as neural networks \citep{YANG2019336} and random forests, to handle interval-valued data based on random set theory, rather than simply applying existing tools to model the interval bounds.
Moreover, other dimension reduction techniques for interval-valued data can be proposed, including new principal component analysis based on $D_K$-distance and interval factor models \citep{He2022Large,wang2012cipca}.

\newpage
{
\baselineskip=16pt
\bibliographystyle{apa}

\bibliography{ref}
}

\newpage
\renewcommand{\thesection}{A}
\setcounter{table}{0}
\renewcommand{\thetable}{A.\arabic{table}}
\setcounter{equation}{0}
\setcounter{remark}{0}
\renewcommand{\theremark}{A.\arabic{remark}}

{\centering\section*{Online Appendix of \\``Sparse Interval-valued Time Series Modeling with Machine Learning"}}
{\centering\subsection*{Appendix A}}
\noindent Appendix A contains the required assumptions.
\begin{assume} \label{ass1}
	$\{Y_t,X_{1,t},...,X_{J,t}\}$ are strictly stationary and ergodic interval stochastic processes with $\mathbb{E}\lVert Y_t\rVert _K^4<\infty$ and $\mathbb{E}\lVert X_{j,t}\rVert _K^4<\infty$ for $j=1,...,J$. The interval innovation $\{u_t\}$ is an IMDS with respect to the information set $\mI_{t-1}$, i.e., $\mathbb{E}(u_t|\mI_{t-1})=[0,0]$ a.s., and $\mE(||u_t||_K^2|\mI_{t-1})=\sigma^2$.
\end{assume}


\begin{assume} \label{ass3}
	(a) The parameter space $\Theta$ is a compact space of $\mathbb{R}^m$; (b) $\btheta^0$ is an interior point in $\Theta$, which is the true parameter vector in ACIX model.
\end{assume}

\begin{assume} \label{ass4}
	$\mE[\langle s_{\frac{\partial u_t(\btheta)}{\partial \btheta}},s'_{\frac{\partial u_t(\btheta)}{\partial \btheta}}\rangle_K]$ and $\mE[\langle s_{\frac{\partial u_t(\btheta)}{\partial \btheta}},s_{u_t(\btheta)}\rangle_K \langle s_{u_t(\btheta)},s_{\frac{\partial u_t(\btheta)}{\partial \btheta}}\rangle_K]$ are positive definite for all $\btheta$ in a small neighborhood of $\btheta^0$, where $u_t(\btheta) = Y_t-\Z_t'\btheta$ and $\frac{\partial u_t(\btheta)}{\partial \btheta} = \frac{\partial (Y_t-\Z_t'\btheta)}{\partial \btheta} = -\Z_t$.
\end{assume}

\begin{assume} \label{ass5}
    The smallest and largest eigenvalues of $\bSigma_T$, denoted as $\rho_{1T}$, $\rho_{2T}$, where $\bSigma_{T}=\frac{1}{T}\sumt\langle s_{\Z_t},s_{\Z_t'}\rangle_K$, satisfy the conditions that $\bSigma_{T}$ is positive definite, and $\rho_{1T}>0$.
\end{assume}

\begin{assume} \label{ass6}
	There exist constants $c_3$ and $c_4$ such that
	\[
	0<c_3\leq\min\{|\theta_{1j}|,1\leq j \leq k_T\}	\leq \max\{|\theta_{1j}|,1\leq j \leq k_T\}	\leq c_4<\infty.
	\]
\end{assume}

\begin{assume} \label{ass7}
There exist constants $\tau_1$ and $\tau_2$ such as $0<\tau_1\leq\tau_{1T}\leq\tau_{2T}\leq\tau_2<\infty$, where $\tau_{1T}$ and $\tau_{2T}$ are the smallest and largest eigenvalues of $\bSigma_{1T} = \frac{1}{T}\sumt\langle s_{\Z_{1t}},s_{\Z_{1t}'}\rangle_K$.
\end{assume}

\begin{assume} \label{ass8}
	(a) $\lambda_T/\sqrt{T}\rightarrow0$, as $T\rightarrow \infty$; (b) $\lambda_TT^{(\gamma-1)/2}\rightarrow\infty$, as $T\rightarrow \infty$, where $\gamma$ a constant given in \eqref{eq2.2}.
\end{assume}

\begin{assume} \label{ass9}
	(a) $\lambda_T(k_T/T)^{1/2}\rightarrow 0$, as $T\rightarrow \infty$;
    (b) $\lambda_T \rho_{1T}^{1+\gamma} T^{(\gamma-1)/2} p^{(-1-\gamma)/2} \rightarrow \infty$, as $T\rightarrow \infty$.
\end{assume}

\begin{assume} \label{ass10}
	$(p+\lambda_Tk_T[\rho_{1T}^{-1}(p/T)^{1/2}+1]^{-\gamma})/(T\rho_{1T}) \rightarrow 0$, as $T\rightarrow \infty$.
\end{assume}

The stationarity and ergodicity of ITS have been defined in \cite{han2016vector}, and the ACIX model is a special form of IVARMA model with one dimension. Assumption \ref{ass1} requires that the fourth moment of $\{Y_t\}$ is finite with respect to $D_K$-metric. $\mE(u_t|\mI_{t-1})=[0,0]$ implies that the support function of the interval innovation $u_t$ is satisfying $\mE(s_{u_t}|\mI_{t-1})=0$. The IMDS assumption of $u_t$ is weaker than i.i.d. condition and weaker than a conditional homoscedastic assumption for the left/right bounds of $u_t$.
Assumption \ref{ass3} specifies the restricted condition of the parameter space and the true parameter. Under Assumption \ref{ass4}, the expectations of matrices $\langle s_{\Z_t},s_{\Z_t'}\rangle_K$ and $\langle s_{\Z_t},s_{u_t}\rangle_K\langle s_{u_t},s_{\Z_t}\rangle_K$ are positive definite, and similar conditions can be found in \cite{fan2004nonconcave}.

Assumption \ref{ass5} implies that $\bSigma_{T}$ is nonsingular for each $T$, and it is worth noting that we allow $\rho_{1T}\rightarrow0$ as $T\rightarrow\infty$. It can be seen in Theorem \ref{thm7} that $\rho_{1T}$ affects the convergence rate of the estimators.
Assumption \ref{ass6} assumes that the nonzero coefficients are uniformly bounded away from zero and infinity, which is the similar to Condition (A4) in \cite{huang2008asymptotic} and Condition (A5) in \cite{huang2008adaptive}.
Assumption \ref{ass7} is a more strict condition than Assumption \ref{ass5}, and it implies that the eigenvalues of $\bSigma_{1T}$ are uniformly bounded. Noticing that $k_T$ is much smaller than $T$ due to the sparsity of predictors, it is reasonable to give this assumption. It is similar to the Condition (A5) in \cite{huang2008asymptotic} for point-valued data.

Assumption \ref{ass8} states that $\lambda_T=o(\sqrt{T})$. When $\gamma=1$, Assumption \ref{ass8}(b) implies $\lambda_T\rightarrow\infty$. 
Assumption \ref{ass9} represents an adaptation of Assumption \ref{ass8} for scenarios where the model's dimensionality diverges. Specifically, Assumption \ref{ass8} can be derived from Assumption \ref{ass9} when $p$ is finite. These two assumptions are essential for establishing the convergence rate in our proofs. Notably, similar conditions have been employed in existing studies, such as the assumptions of Theorem 2 in \cite{2001Variable}, the assumptions of Theorem 2 in \cite{zou2006adaptive}, and Conditions (A2)-(A3) in \cite{huang2008asymptotic}.
Assumption \ref{ass10} makes further assumptions about the relationships between parameters to ensure the consistency of the penalized minimum distance estimation when $p$ is diverging.


\newpage
\setcounter{equation}{0}
\renewcommand{\theequation}{A.\arabic{equation}}
\renewcommand{\thesubsection}{A.\arabic{subsection}}
\baselineskip=18pt

\end{document}